\documentclass{aa}
\usepackage[switch]{lineno}
\modulolinenumbers[10]
\usepackage[T1]{fontenc}
\usepackage{amsmath} 
\usepackage{amssymb}
\usepackage{amstext}
\usepackage{booktabs}
\usepackage{graphicx}
\usepackage{yfonts}
\usepackage{xcolor}
\usepackage{txfonts}
\usepackage{subfigure}
\usepackage{float}
\usepackage{natbib,twoopt}
\usepackage{xspace}
\usepackage{balance}
\usepackage{upgreek}
\usepackage{multirow}
\usepackage{multicol}
\usepackage[colorlinks=true, allcolors=blue]{hyperref}
\usepackage[normalem]{ulem}
\usepackage{enumitem}
\usepackage{soul}
\usepackage[normalem]{ulem}
\usepackage{siunitx}
\usepackage[flushleft]{threeparttable}
\usepackage[nameinlink, capitalise]{cleveref}
\usepackage{orcidlink}
\usepackage{bm}
\usepackage{lastpage}

\def\apjl{ApJL}
\def\apjs{ApJS}

\def\aap{A\&A}

\newcommand{\vect}[1]{\boldsymbol{#1}}

\newcommand{\ii}{\mathrm{i}}

\newcommand{\vtheta}{\vect{\theta}}
\newcommand{\vbeta}{\vect{\beta}}

\newcommand{\zD}{z_{\mathrm{l}}}
\newcommand{\zS}{z_{\mathrm{s}}}
\newcommand{\chiD}{\chi_{\mathrm{l}}}
\newcommand{\chiS}{\chi_{\mathrm{s}}}
\newcommand{\fK}{f_{K}}
\newcommand{\fD}{f_{\mathrm{l}}}
\newcommand{\fS}{f_{\mathrm{s}}}
\newcommand{\fDS}{f_{\mathrm{ls}}}

\newcommand{\mapsimm}{\textsc{MapSim}}
\newcommand{\mapsim}{\textsc{MapSim} }
\newcommand{\lcdm}{$\mathrm{\Lambda CDM}$ }

\newcommand{\dustp}{{\small DUSTGRAIN}-\emph{pathfinder} }

\newcommand\lsim{\mathrel{\rlap{\lower4pt\hbox{\hskip1pt$\sim$}}
        \raise1pt\hbox{$<$}}}
\newcommand\gsim{\mathrel{\rlap{\lower4pt\hbox{\hskip1pt$\sim$}}
        \raise1pt\hbox{$>$}}}

\bibpunct{(}{)}{;}{a}{}{,} 
\DeclareGraphicsExtensions{.eps,.ps,.pdf}
\begin{document}
\title{Weak-lensing tunnel voids in simulated light cones: A new pipeline to investigate modified gravity and massive neutrinos signatures}

\author{
Leonardo Maggiore\inst{\ref{1},\thanks{\email{\href{mailto:leonardo.maggiore3@unibo.it}{leonardo.maggiore3@unibo.it}}}}, Sofia Contarini\inst{\ref{2}}, Carlo Giocoli\inst{\ref{3},\ref{4}}, Lauro Moscardini\inst{\ref{1},\ref{3},\ref{4}}}

\institute{
Dipartimento di Fisica e Astronomia "Augusto Righi" - Alma Mater Studiorum Universit\`{a} di Bologna, via Piero Gobetti 93/2, 40129 Bologna, Italy
\label{1} \and
Max Planck Institute for Extraterrestrial Physics, Gießenbachstraße 1, 85748 Garching, Bayern, Germany
\label{2} \and
INAF-Osservatorio di Astrofisica e Scienza dello Spazio di Bologna, Via Piero Gobetti 93/3, 40129 Bologna, Italy
\label{3} \and
INFN-Sezione di Bologna, Viale Berti Pichat 6/2, 40127 Bologna, Italy
\label{4}
}

\idline{A\&A, 701, A55 (2025)}
\doi{\href{https://doi.org/10.1051/0004-6361/202554968 }{https://doi.org/10.1051/0004-6361/202554968 }}
\AANum{A55}
\yearCop={2025}

\date{Received 1 April 2025 / Accepted 6 July 2025}

\abstract{Cosmic voids offer a unique opportunity to explore modified gravity (MG) models. Their low-density nature and vast extent make them especially sensitive to cosmological scenarios of the class $f(R)$, which incorporate screening mechanisms in dense, compact regions. Weak lensing (WL) by voids, in particular, provides a direct probe for testing MG scenarios. While traditional voids are identified from 3D galaxy positions, 2D voids detected in WL maps trace underdense regions along the line of sight and are sensitive to unbiased matter distribution. To investigate this, we developed an efficient pipeline for identifying and analyzing tunnel voids, namely, underdensities detected in WL maps, specifically in the signal-to-noise ratio (S/N) of the convergence. In this work, we used this pipeline to generate realistic S/N maps from cosmological simulations featuring different $f(R)$ scenarios and massive neutrinos, comparing their effects against the standard $\Lambda$CDM model. Using the convergence maps and the 2D void catalogs, we analyzed various statistics, including the probability density function, angular power spectrum, and void size function. We then focused on the tangential shear profile around 2D voids, demonstrating how the proposed void-finding algorithm maximizes the signal. We show that MG leads to deeper void shear profiles due to the enhanced evolution of cosmic structures, while massive neutrinos have the opposite effect. Furthermore, we find that parametric functions typically applied to 3D void density profiles are not suitable for deriving the shear profiles of tunnel voids. Therefore, we propose a new parametric formula that provides an excellent fit to the void shear profiles across different void sizes and cosmological models. Finally, we test the sensitivity of the free parameters of this new formula to the cosmological model, revealing its potential as a probe for detecting the effects of MG models and the presence of massive neutrinos.}


\authorrunning{Maggiore, L., et al.}
\titlerunning{A\&A, 701, A55 (2025)}
\maketitle
\makeatletter
\gdef\aa@idline{}
\gdef\aa@doi{}
\makeatother

\section{Introduction}
Despite the success of the $\Lambda$-cold dark matter ($\Lambda$CDM) concordance model \citep{Heavens2017}, which accurately describes observations from well-tested Solar System dynamics \citep{Will2014, Baker2021} to gravitational wave propagation \citep{Creminelli2017, LIGO2017}, some tension remains among the observations. In particular, there are discrepancies between cosmic microwave background (CMB) \citep{Planck2020a, Planck2020b} constraints and the lensing measurements of $\sigma_8$ \citep{Abdalla2022, DiValentino2021_sigma8} as well as the supernovae determinations of $H_0$ \citep{DiValentino_2021_Hubble}. These discrepancies have motivated tests of gravity on cosmic scales \citep{Koyama2016}.

In response, modified gravity (MG) theories have been proposed to address observational tensions \citep{Carrol2001, Martin2012, Moresco&Marulli2017}, including the possibility that general relativity (GR) fails on cosmological scales \citep{Dolgov2003, Clifton2012, Joyce2015, Ishak2019}. These models introduce additional degrees of freedom, modifying structure formation and spacetime geometry. Among them, $f(R)$ gravity mimics dark energy (DE) effects via an additional scalar field while employing screening mechanisms to restore GR predictions at small scales and in high-density regions \citep{Joyce2015, Bertotti2003, Will2005, Hinterbichler2010, BraxValageas2013MG}. Testing MG models is challenging due to their similarity to $\Lambda$CDM, but cosmic voids offer a promising avenue \citep{Cautun2014}. Their low densities weaken screening, making them ideal for detecting deviations from GR \citep{Barreiraetal2015, Baker2018}.\\
\indent A powerful approach to studying voids is through the weak lensing (WL) phenomenon, which probes the deflection of photons by large-scale structure (LSS). It introduces distinct distortions in background galaxies, quantified by cosmic shear $\gamma$ and convergence $\kappa$ \citep{Bartelmann&Schneider2001WeakLensing, Kilbinger2015ReviewCosmologyCosmicShearObservations, Ishak2019, Umetsu20} and is highly sensitive to the growth of structure and cosmic expansion \citep{Bartelmann2010REVIEWLensing, Troxel&Ishak2015WeakLensingEraPrecisionCosmology}. The next generation of Stage-IV WL surveys, including Rubin \citep{LSST}, \textit{Euclid} \citep{Laureijs11, EC_higherorderWL_Ajani2023, EC_Congedo2024_WL}, and Roman \citep{Roman2015}, will significantly improve sky coverage and angular resolution, enhancing constraining power. Euclid DR1 \citep{EC_Mellier_Overview2024}, in particular, will increase WL source counts by a factor of 4-5, extending the redshift distribution \citep{Kitching2017}, with respect to ground-based surveys like KiDS \citep{KIDS2017, KiDS2023}. To fully exploit these data, accurate numerical simulations and high-fidelity modeling techniques are essential.

In voids, the deflection of light occurs outward, opposite to the inward deflection observed in overdense areas. This phenomenon is due to the gravitational effects of spacetime curvature, driven by matter overdensities, which influence the trajectory of light. Additionally, the magnitudes of observed astronomical objects are influenced by the environment and line of sight inhomogeneities, leading to magnification or demagnification effects \citep[see also][]{Clarkson2012, Bolejko2012}. The unique reversal of gravitational lensing features, such as shear and convergence profiles, is commonly referred to as anti-lensing \citep{Bolejko2013_Antilensing}.

\cite{Cautun2018} compared the predictions of 3D and 2D voids for \textit{Euclid} and LSST-like lensing surveys, finding that 2D voids are more sensitive to MG theories, showing a deeper lensing signal due to enhanced growth of cosmic structures. Moreover, WL studies applied to voids offer an advantage by reducing the reliance on luminous tracers needed for 3D void identification. In fact, WL is sensitive to the total matter distribution, enabling the detection of voids directly in the matter domain. Two primary methodologies are commonly employed in the literature for analyzing WL around 2D voids:
\begin{itemize}
    \item WL tunnel voids: This approach measures tangential shear ($\gamma_t$) or convergence ($\kappa$), which reflect the distortions of background galaxy shapes caused by the projected profiles along the line of sight of extensive underdense structures, typically spanning hundreds of megaparsecs \citep{Higuchi2013, Gruen2015, Davies2021ConstrainingCosmoWeakLensingVoids, Shimasue2024}.
    \item Void lensing (VL): This method quantifies the excess surface mass density, representing the projection of total matter within a thin lens. This technique is suitable for voids with radii $\leq 50 \ h^{-1}$ Mpc, where the thin lens approximation holds, and corresponds to studying the WL signal of voids using a tomographic approach \citep[see][for further details]{Boschetti2023}.
\end{itemize}
The first method provides less 3D information, but primarily leverages tangential shear for constraining alternative gravity models, as shown by \cite{Cautun2018} and \cite{ Davies2019CosmologicalTestGravityWeakLensing}. In contrast, VL seeks to retrieve dynamic and morphological information about voids, aligning more closely with their standard 3D definitions. However, tunnels exhibit the highest signal-to-noise ratio (S/N) of $\gamma_t$ in WL maps, producing shear profiles up to ten times stronger \citep{Davies2018Weaklensingbyvoid} and offering superior constraints on MG models due to their small size, high abundance, and minimal overlap \citep{Cautun2018, Davies2019CosmologicalTestGravityWeakLensing, Davies2021OptimalVoidFinderinWeakLensing}. For these reasons, in this work we decided to adopt the WL tunnel approach.

This paper is organized as follows. In Sect. \ref{sec:theory}, we give a theoretical overview of weak gravitational lensing and $f(R)$ gravity models. In Sect. \ref{sec:methods}, we describe the cosmological simulations employed in this work, followed by the construction of WL light cones and the implementation of ray-tracing techniques to generate convergence maps. In Sect. \ref{sec:void_finder}, we present a new code to identify WL tunnel voids, which we applied to the generated convergence maps. Next, in Sect. \ref{sec:statistics}, we introduce and analyze the two primary void statistics explored in this work: the tunnel void size function and the stacked tunnel void tangential shear. In the final part of the paper, Sect. \ref{Sec:modeling}, we describe how we adopted two different approaches to model the tangential shear signal. Specifically, we first integrated two established 3D void surface density profiles along the line of sight and then  introduced a new parametric model to reproduce the tangential shear signal. Finally, in Sect. \ref{sec:conclusions}, we summarize our conclusions.

\section{Theory}\label{sec:theory}
In this section, we introduce the theoretical background of WL and $f(R)$ MG models to describe the fundamental quantities utilized in this work.

\subsection{Weak lensing formalism}\label{subsec:wlform}
The phenomenon of gravitational lensing occurs when light rays from distant sources pass through inhomogeneous matter distributions, experiencing deflections that distort the observed shapes of background sources \citep{Bartelmann&Schneider2001WeakLensing, Kilbinger2015ReviewCosmologyCosmicShearObservations}. To model this phenomenon, it is assumed that on scales relevant to gravitational light deflection, spacetime is described by a weakly perturbed Friedmann-Lemaître-Robertson-Walker metric.

Assuming the Newtonian gauge and defining $\Phi$ and $\Psi$ as the Einstein-frame metric potentials, the perturbed metric is described as:
\begin{equation}\label{pert_metric}
\mathrm{d} s^{2}=(1+2 \Psi) \mathrm{d} t^{2}-a^{2}(t)(1-2 \Phi) \mathrm{d} x_{i} \mathrm{~d} x_{j} \, .
\end{equation}
Here, $\Psi$ represents the gravitational potential, which describes the temporal part of the metric perturbations, while $\Phi$ is the gravitational curvature potential, describing the spatial part of the metric perturbations.

Here, we let \(\bm{\beta}\) represent the true angular position of a source at a comoving line of sight distance \(\chiS\) and redshift \(\zS = z(\chiS)\), with \(\bm{\theta}\) as its observed angular position on the sky. In this framework, the angular positions \(\bm{\beta}\) and \(\bm{\theta}\) are connected via the lens equation, which traces the trajectory of the photon as it is deflected by an angle \(\bm{\alpha}\) through a lens or lens system. It is generalized as: 
\begin{equation}
\label{eq:lens_equation}
\bm{\beta} (\bm{\theta}, \zS) = \bm{\theta} - \bm{\alpha} 
= \bm{\theta} - \frac{2}{c^2} \int_0^{\chiS} \frac{\fDS}{\fD \fS}
\nabla_{\bm{\beta}} \Phi_{\mathrm{len}}\bigl(\bm{\beta} (\bm{\theta}, \chiD), \chiD, \zD \bigr) \, \mathrm{d}\chiD \, ,
\end{equation}
where \(c\) is the speed of light and \(\fDS = \fK(\chiS - \chiD)\), \(\fD = \fK(\chiD)\), \(\fS = \fK(\chiS)\), where we assume the case $\chiS > \chiD$ with ``s'' and ``l'' representing the source and the lens, respectively. Here, we assume a generic curvature \( K \), so \(\fK(\chi)\) represents the comoving angular diameter distance, and \(\fK(\chi)\bm{\theta}\) corresponds to the transverse comoving vector for a comoving line-of-sight distance \(\chi=\chi(z)\) \citep{Misner1973, cosmodistances1999, Umetsu20}. $\Phi_{\mathrm{len}}\bigl(\vbeta, \chiD, \zD \bigr)$ denotes the 3D lensing gravitational potential generated by a mass distribution $\rho(\vbeta, \chiD)$ at redshift $\zD$ \citep{Hilbert2020}. 

Generally, $\Phi_{\rm len}=(\Phi + \Psi)/2$, but if we also assume the negligibility of anisotropic stress in the perturbed metric, as is often the case for non-relativistic matter (e.g., cold dark matter), the two metric potentials become equal. Thus, we have:
\begin{equation}\label{eq:Newtonian_potential}
\Phi = \Psi = \Psi_{\rm N} \, ,
\end{equation}
where $\Psi_{\rm N}$ is the Newtonian potential. This equality holds in the standard GR model, where the accelerated expansion of the Universe is driven by a cosmological constant, so we can use $\Phi_{\rm len}=\Phi$ under this assumption. $\Phi$ is related to the non-relativistic matter density contrast, $\delta_m=\delta\rho/\bar{\rho}$, through the Poisson equation:
\begin{equation}\label{eq:Poisson_equation}
\nabla^2\Phi\bigl(\bm{\beta}, \chiD, \zD \bigr) = 4 \pi G a^2 \delta\rho(\bm{\theta}, \chiD, \zD) = \frac{3 \Omega_m H_0^2}{2 a} \delta_m(\bm{\theta}, \chiD, \zD) \, ,
\end{equation}
where $G$ is the gravitational constant, $\bar{\rho}$ is the mean matter density of the Universe, $\Omega_m$ is the total matter density parameter, and $H_0$ represents the Hubble parameter at the present time.

The image distortion is described by the Jacobian distortion matrix $\mathbf{A}(\bm{\theta}, \chiS)$, which represents the relative positions of nearby light rays and is obtained by differentiating the lens equation with respect to $\bm{\theta}$. Assuming the flat-sky approximation, where $\bm{\beta}$ and $\bm{\theta}$ are 2D Cartesian coordinate vectors \citep{Becker2013}, we can express it as
\begin{equation}
    A_{ij}(\bm{\theta}, \chiS) \equiv \frac{\partial \beta_{i} (\bm{\theta}, \zS)}{\partial \theta_{j}} = \delta_{ij} - \frac{\partial \alpha_{i}(\bm{\theta}, \zS, \zD)}{\partial \theta_{j}} \, ,
    \label{eq:jacobian}
\end{equation}
and the distortion matrix $\mathbf{A}(\bm{\theta}, \chiS)$ takes the form:
\begin{equation}\label{eq:amp_mat}
        \mathbf{A}(\bm{\theta}, \chiS) \approx
        \begin{pmatrix}
            1 - \kappa - \gamma_1 & -\gamma_2 - \omega \\
            -\gamma_2 + \omega & 1 - \kappa + \gamma_1 \\
        \end{pmatrix} \, ,
\end{equation}
where $\delta_{ij}$ is the Kronecker delta. This defines the lensing convergence, $\kappa$, the lensing complex shear, $\gamma = \gamma_1 + \ii \gamma_2$, and the lensing rotation, $\omega$, which is $\mathcal{O}(\Phi^2)$ \citep{Petri2017}. In the WL regime, we have $ |\kappa|, |\gamma|, |\omega| \ll 1 $.

To solve the integral form of Eq.~\eqref{eq:jacobian}, a common approach is to expand it in a series of $\Phi$, retaining only the first-order term, known as the Born approximation \citep{Bartelmann&Schneider2001WeakLensing}. This approximation assumes negligible rotation and disregards lens-lens coupling, allowing the deflection angle $\bm{\alpha}$ to be expressed as the gradient of a 2D lensing potential, $\psi_{\rm len}$. Under this framework, the WL signal arises from the cumulative matter distribution along the line of sight \citep{Davies2021OptimalVoidFinderinWeakLensing}. The convergence, $\kappa$, follows from the 2D Poisson equation and can be written in terms of the matter density contrast, $\delta_{\rm m}$, as:
\begin{equation}\label{eq:kappa_born}
\kappa_{\rm Born}(\bm{\theta}, \zS) = \frac{3 H_0^2 \Omega_m}{2 c^2} \int^{\chiS}_0 (1+\zD) W(\chiD,\chiS) \delta_{\rm m} (\bm{\theta}, \chiD, \zD) \, \mathrm{d}\chiD \, ,
\end{equation}
where $W(\chiD,\chiS) \equiv (\fD \fDS)/\fS$ is defined as the lensing kernel or lensing efficiency factor. This demonstrates that the measured WL convergence can be understood as the integration of the matter density along the unperturbed line of sight, weighted by the lensing kernel.

Moreover, the radial convergence profile of an object \( \kappa(r_p) \) is related to its radial tangential shear profile through the equation \citep{Davies2018Weaklensingbyvoid, Davies2021OptimalVoidFinderinWeakLensing}:
\begin{equation}\label{eq:gamma_t}
\gamma_{\rm{t}}(r_p) = \bar{\kappa}(< r_p) - \kappa (r_p) \, ,
\end{equation}
where $\bar{\kappa}(< r_p)$ is the mean enclosed convergence within the radius, \( r_p \), defined as
\begin{equation}\label{eq:kappa_enc}
\bar{\kappa}(< r_p) = \frac{1}{\pi r_p^2}\int_0^{r_p} 2 \pi r_p' \kappa(r_p') dr_p' \, .
\end{equation}
Based on the assumption of the flat-sky approximation, here and throughout this work, \( r_p \) is used to denote the projected distance from the void center instead of \( \theta \).

\subsection{$f(R)$ modified gravity models}\label{f(R)}
In this work, we analyzed a class of models called $f(R)$ \citep{Weyl1918}, which is a set of alternative fourth-order scalar-tensor theories of gravity. These theories have been extensively investigated as models deviating from GR \citep[e.g.,][]{Utiyama1962, Starobinsky1980} due to their potential to explain both the early and late-time acceleration of the Universe \citep{Starobinsky2007, HuSawiki2007FR, Nojiri2008_a, Nojiri2011}. Specifically, they mimic the observed accelerated expansion of the Universe by introducing an additional scalar field that acts as a fifth force, repulsive on large scales. This scalar field replaces the cosmological constant $\Lambda$ of the standard $\Lambda$CDM model and differs from GR in the evolution of density perturbations due to gravitational instability. They also use screening mechanisms to avoid Solar System constraints, aligning with well-tested GR predictions on small scales and in overdense regions.

$f(R)$ theories of gravity extend the Einstein-Hilbert action by incorporating functions of the Ricci scalar, $R$, which can result in fourth-order field equations unless a constant term is added to the gravitational Lagrangian \citep{Buchdahl1970}. The action for $f(R)$ gravity is given by
\begin{equation}\label{eq:fr_action}
S=\frac{M_{\mathrm{P}}^{2}}{2} \int \mathrm{d}^{4} x \sqrt{-g}(R+f(R))+S_{m}\left[g_{\mu \nu}, \psi\right],
\end{equation}
where $M_{\mathrm{P}}^{2}=1/(8 \pi G)$ is the reduced Planck mass and $S_{m}$ is the action of the matter field, $\psi$. The function $f(R)$ controls deviations from GR, becoming significant in the low curvature regime ($R \rightarrow 0$) and in low-density environments such as voids \citep{Brans1961}. GR can be recovered by setting $f = -2 \Lambda^{\mathrm{GR}}$ for a more generalized cosmic acceleration, as given in \cite{Carrol2004}.

A prominent $f(R)$ model proposed by \cite{HuSawiki2007FR} is expressed as
\begin{equation}
f(R)=-m^{2} \frac{c_{1}\left(\frac{R}{m^{2}}\right)^{n}}{c_{2}\left(\frac{R}{m^{2}}\right)^{n}+1},
\end{equation}
where $m^{2} \equiv H_{0}^{2} \Omega_{\mathrm{m}}$ defines the mass scale, while $c_{1}, c_{2},$ and $n$ are non-negative parameters. For $\Lambda$CDM consistency, the condition $c_{1}/c_{2} = 6 \Omega_{\Lambda}/\Omega_{\mathrm{m}}$ must hold. In the limit $c_{2}\left(R / m^{2}\right)^{n} \gg 1$, the scalar field, $f_{R} \equiv \mathrm{d} f(R) / \mathrm{d} R$, can be approximated as
\begin{equation}
f_{R} \approx -n \frac{c_{1}}{c_{2}^{2}}\left(\frac{m^{2}}{R}\right)^{n+1}.
\end{equation}
For $n=1$, the model simplifies, with $f_{R 0}$ defined as:
\begin{equation}\label{eq:fr0}
f_{R 0} \equiv -\frac{1}{c_{2}} \frac{6 \Omega_{\Lambda}}{\Omega_{\mathrm{m}}}\left(\frac{m^{2}}{R_{0}}\right)^{2},
\end{equation}
where $R_{0}$ is the background value of $R$ at the present time.

By modifying the action in Eq. \eqref{eq:fr_action} with respect to $g_{\mu \nu}$, we obtain the modified Einstein equations for $f_{R}$ as
\begin{equation}
f_R R_{\mu\nu} - \frac{1}{2}f g_{\mu\nu} - \nabla_\mu \nabla_\nu f_R + g_{\mu\nu} \Box f_R = 8\pi G T_{\mu\nu} \, ,
\end{equation}
where \(\nabla\) is the covariant derivative and \(\Box \equiv g^{\mu\nu} \nabla_\mu \nabla_\nu\) is the D'Alembert operator. From its trace, we derive the motion equation for the scalar field:
\begin{equation}\label{eq:motion_fR}
\nabla^{2} \delta f_{R}=\frac{a^{2}}{3}\left[\delta R\left(f_{R}\right)-8 \pi G \delta \rho\right].
\end{equation}
By extracting its time-time component and assuming small perturbations $\delta f_{R}$, $\delta R$, and $\delta \rho$ on a homogeneous background and the quasi-static field approximation (slow variation for $f_{R}$), we obtain the equivalent of the Poisson equation for scalar metric perturbations, $2 \psi=\delta g_{00} / g_{00}$, as
\begin{equation}\label{eq:poisson_fR}
\nabla^{2} \psi=\frac{16 \pi G}{3} a^{2} \rho-\frac{a^{2}}{6} \delta R\left(f_{R}\right).
\end{equation}
Combining Eqs.\ \eqref{eq:motion_fR} and \eqref{eq:poisson_fR} allows  us to derive exact solutions for extreme cases and analyze the scale and manner in which $f(R)$ gravity diverges from GR.

Within this framework, the behavior of $f(R)$ models varies depending on $f_{R 0}$ values relative to the effective potential, the Newtonian potential, $\Psi_{N}$. If $f_{R 0}\ll\Psi_{N}$, the behavior of the model closely recovers GR in high-curvature regions due to the screening mechanism. The most notable example of these mechanisms is the so-called Chameleon screening mechanism \citep{Khoury2004_ChameleonCosmo, Khoury2004_ChameleonFields}. The response of the scalar field to the effective potential, influenced by external matter sources, varies with density, increasing its effective mass in high-density regions and decreasing it in low-density ones (Chameleon field). Conversely, if $f_{R 0}\gg\Psi_{N}$, gravity would be over-amplified. Therefore, $f_{R 0}$ should be close in magnitude to $\Psi_{N}$, typically  between $10^{-5}$ and $10^{-6}$. A value of $10^{-4}$ may still be acceptable when including the contribution of other components like massive neutrinos. In fact, it has been shown that according to the mass associated with these particles, the effect caused by LSS can be nearly counteractive to $f(R)$ models. 
This happens because, for $\left|f_{R 0}\right| \gg|\psi|$, the scale size of the scalar field is comparable to the scale of neutrino thermal free-streaming. This creates a degeneracy visible in the matter power spectrum \citep{Saito2008, Wagner2012, Cataneo2015}, the halo mass function \citep{Marulli2011, Villaescusa2013}, the clustering properties of CDM halos and redshift-space distortions \citep{Zennaro2018, Garcia-Farieta2019}. We note that a possible hint towards the disentangling of the combined effects of $f(R)$ models and massive neutrinos was found analyzing the abundance of large voids at high redshifts \citep{Contarini2021_voids_MG_neutrinos}. We  investigate the impact of these separate components using WL tunnel voids in Sect. \ref{sec:statistics}. 

We go on to focus on the effect of MG on the WL phenomenon. Using the perturbed metric in Eq. \eqref{pert_metric}, WL offers a direct connection to gravity theories through the 3D lensing potential $\Phi_{\mathrm{len}}$ described in Sect. \ref{subsec:wlform}, because it is defined as the sum of the two metric potentials. In contrast to what happens in GR (see Eq. \ref{eq:Newtonian_potential}), differentiating these potentials is often crucial for analyzing models, as the additional scalar field may affect photons and matter differently.

In $f(R)$ models, using the general definition of $\Phi_{\mathrm{len}}$ in Eq. \eqref{eq:Poisson_equation}, the two modified Poisson equations for the metric potentials can be derived from the variation of the Einstein equations as
\begin{equation}\label{eq:metric_poiss_eq}
\nabla^{2} \Phi = 4 \pi G a^{2} \delta \rho - \frac{c^{2} \nabla^{2}}{2} \delta f_{R} , \, \, \, \, \nabla^{2} \Psi = 4 \pi G a^{2} \delta \rho + \frac{c^{2} \nabla^{2}}{2} \delta f_{R} \, .
\end{equation}
In terms of the Newtonian potential, these can be expressed as:
\begin{equation}
\Phi = \Psi_{N} - \frac{c^{2}}{2} \delta f_{R}, \quad \Psi = \Psi_{N} + \frac{c^{2}}{2} \delta f_{R} \, .
\end{equation}
Consequently, the lensing potential remains unaffected as the corrections cancel out, preserving the standard GR form \citep{HuSawiki2007FR, Giocolietal2018_mio}. We conclude that in $f(R)$ models, deviations from GR arise not from changes in the lensing potential $\Phi_{\mathrm{len}}$, but only from alterations in the LSS matter distribution. Therefore, we would expect a different distribution of convergence and shear compared to GR predictions. In particular, in regions of low density, the scalar field $\delta f_R$ contributes to an enhancement of the gravitational interaction, resulting in a more pronounced WL signal around voids than in $\Lambda$CDM models.

\section{Methods}\label{sec:methods}
\subsection{DUSTGRAIN-\textit{pathfinder} simulations}\label{DUST-path}
In our analysis, we used data produced in \cite{Giocolietal2018_mio} from the Dark Universe Simulations to Test Gravity In the Presence of Neutrinos (DUSTGRAIN) project, a suite of N-body cosmological simulations.

The DUSTGRAIN-\textit{pathfinder} runs are cosmological collisionless simulations that track the evolution of $768^{3}$ cold dark matter (CDM) particles within a periodic box with a side length of $750 \, h^{-1} \, \mathrm{Mpc}$. These particles have a mass of $m^{p}_{\mathrm{cdm}}=8.1 \times 10^{10} \, h^{-1} M_{\odot}$ and their gravitational softening has been set to $\varepsilon_{g}=25 \, h^{-1} \mathrm{kpc}$, corresponding approximately to $1/40$th of the mean inter-particle separation. All simulations assume a constant $\Omega_{\mathrm{m}}$ to ensure matching power spectra on large scales, with and without neutrinos. The reference $\Lambda$CDM simulation assumes GR, $M_{\nu}=0 \, \mathrm{eV}$. Its cosmological parameters follow Planck2015 constraints \citep{Planck2016a_cosm_params}, with $\Omega_{\mathrm{m}}=\Omega_{\mathrm{cdm}}+\Omega_{\mathrm{b}}+\Omega_{\nu}=0.31345$, $\Omega_{\Lambda}=0.68655$, $H_{0}=67.31 \, \mathrm{km} \, \mathrm{s}^{-1} \, \mathrm{Mpc}^{-1}$, $n_{s}=0.9658$, and $\sigma_{8}=0.842$ at $z=0$.
\begin{table*}
\centering
\caption{Cosmological models employed}
\renewcommand{\arraystretch}{1.3}
\begin{tabular}{lcccccc}
\hline
Simulation & Gravity & $f_{R0} $ & $M_{\nu }$ [eV] & $\Omega_{\mathrm{cdm}}$ & $\Omega_{\nu }$ & $m^{p}_{\mathrm{cdm}} \ [h^{-1} \ M_{\odot}]$ \\
\hline
$\Lambda$CDM & GR & - & 0 & 0.31345 & 0 & $8.1\times 10^{10}$ \\
$fR4$ & $f(R)$ & $-1\times 10^{-4}$ & 0 & 0.31345 & 0 & $8.1\times 10^{10}$ \\
$fR5$ & $f(R)$ & $-1\times 10^{-5}$ & 0 & 0.31345 & 0 & $8.1\times 10^{10}$ \\
$fR6$ & $f(R)$ & $-1\times 10^{-6}$ & 0 & 0.31345 & 0 & $8.1\times 10^{10}$ \\
$\Lambda$CDM$_{0.15}$ & GR & - & 0.15 & 0.30987 & 0.00358 & $8.01\times 10^{10}$ \\
\hline
\end{tabular}
\tablefoot{Summary of the main numerical and cosmological parameters of the subset of the \dustp simulations considered in this work. Here, $f_{R0}$ represents the MG parameter, $M_\nu$ the neutrino mass, $m^p_{\rm cdm}$ the CDM particle mass, and $\Omega_{\rm cdm}$ and $\Omega_{\nu}$ the $\mathrm{CDM}$ and neutrino density parameters, respectively.}
\label{tab_fr}
\end{table*}
\indent The alternative cosmological models included in this simulation suite are characterized by deviations from GR in the form of MG of class $f(R)$. The strength of the scalar field is regulated by the parameter $fR0$, which is set to $10^{-4}$, $10^{-5}$ or $10^{-6}$. These simulations are referred to as $fR4$, $fR5$, and $fR6$, respectively.

Additionally, cosmological scenarios featuring massive neutrinos are also included in the DUSTGRAIN-\textit{pathfinder} suite, with the goal of exploring the degeneracies between the effects of this hot dark matter component and $f(R)$ gravity. However, in this paper, we want to focus on the isolated signature of neutrinos, leaving the more complex task of disentangling competing effects for future work. For this purpose, we consider a standard $\Lambda$CDM simulation with neutrinos of a mass $M_\nu = 0.15 \ \mathrm{eV}$, which we refer to as $\Lambda$CDM$_{0.15eV}$. In this case, we tracked the evolution of $2 \times 768^3$ particles considering $m^p_{\rm cdm}=8.01 \times 10^{10}\,h^{-1}M_{\odot}$ and $m^p_{\nu}=9.25\times 10^8\,h^{-1}M_{\odot}$ as dark matter and massive neutrino particle masses, respectively.

The initial conditions of the DUSTGRAIN-\textit{pathfinder} simulations were generated using transfer functions calculated by the \texttt{CAMB} Boltzmann solver \citep{Lewis2000} at the initial redshift $z_{i}=99$. We clarify that they matched across all cosmological models, namely, they share identical random phases in their initial density fields. In the case of simulations including massive neutrinos, two correlated Gaussian random fields were used to initialize CDM and neutrino components from the same initial phase realization, following the methods described in \cite{Zennaro2018, Villaescusa2018}. This setup was deliberately chosen to ensure consistency in the large-scale modes and to allow a direct and faithful comparison across cosmologies, under controlled variance conditions.

To summarize, this work employed five simulation sets: a $\Lambda$CDM reference model, three MG scenarios with different $fR0$ values, and a standard cosmology featuring massive neutrinos. The main characteristics of these simulations are presented in Table \ref{tab_fr}. During each simulation, a sequence of $34$ full comoving snapshots was stored, each representing the specified comoving volume at a particular cosmological epoch.

\subsection{Weak lensing light cones}\label{subsec:LC}
The construction of past light cones is crucial for accurately simulating WL effects produced by the total projected matter density distribution. In this work, we used a post-processing reconstruction method, assuming the flat-sky approximation to slice particle snapshots by their comoving distances from the observer \citep[see][]{Shirasaki2017}.
Specifically, we processed each DUSTGRAIN-\textit{pathfinder} simulation set using the \mapsim routine \citep{Giocolietal2014W&SLensing, Giocolietal2017}, which operates in two main steps, known as i-\mapsim and ray-\mapsimm. 

In the initial phase, i-\mapsimm{}, particle positions from simulation snapshots within the specified field of view (FOV) are projected onto various lens planes along the line of sight. We stacked $21$ of the $34$ available snapshots from $z_{\mathrm{min}}=0$ to $z_{\mathrm{max}}=4$. The light cones have a square base with sides of $5 \, \mathrm{deg}$, corresponding to the angular size of the simulation box at $z_{\mathrm{max}}$, and cover a total sky area of $25 \, \mathrm{deg}^2$. Each light cone is shaped as a pyramid, with the observer at the vertex at $z=z_{min}=0$ and the square base at the comoving distance $\chi(z_{\mathrm{max}})$, as shown in Fig. \ref{fig:LC}. To achieve high-resolution maps of the projected matter density, the stacked snapshots were divided into $27$ lens planes. Since gravitational lensing depends on the projected matter density along the line of sight, each particle was assigned to the closest lens plane based on its moving distances, preserving its angular position within the specified FOV aperture. Simulations with massive neutrinos incorporate this extra component consistently.

The mass density was then interpolated from these projected particle positions to a 2D grid using a triangular-shaped cloud scheme. The grid pixels are designed to have the same angular size across all lens planes. The angular surface mass density $\Sigma(\vtheta)$ on the $l$-th lens plane perpendicular to the $(0,0,1)$ direction, at a pixel with coordinate indices $(i,j)$ is computed as 
\begin{equation}
\Sigma^l_{(i, j)}(\vtheta) = \frac{1}{A_{pix}^l} \sum_{k=1}^{n} m_k \, ,
\end{equation}
where $A_{pix}^l=4\pi/N_{pix}$ is the comoving pixel area in steradians for the $l$-th lens plane, and $m_k$ represents the masses of particles within the pixel, satisfying $f_K(\chi_l) < f_K(\chi_{l'}), \quad \forall l' \neq l \quad \text{and} \quad \chi_k < \chi_s$. In this way, a discretization of the 3D density distribution in several mass maps can be obtained up to the selected source plane $z_{s}$. The construction of light cones from simulations and the projection of the mass distribution onto individual planes was done at the same time by the \mapsimm{} tool in this first step.

To enhance the WL statistics, we generated $256$ semi-independent light-cone realizations for each cosmological model. This was achieved by randomizing the comoving simulation boxes across the redshift range \( z_{\text{obs}} < z < z_{s} \) by applying the following transformations:\\
\\
\begin{itemize}
    \item inverting the sign of the cartesian coordinates;
    \item translating the position of the observer within the box;
    \item permuting the coordinate axes.
\end{itemize}
This procedure preserves the clustering properties of the particle distribution in a snapshot, ensuring the statistical validity of the light cones \citep{Roncarelli2007}. \mapsim algorithm enables the storage of halo and sub-halo catalogs associated with each randomization \citep{Castroetal2018}. The constructed light cones utilize approximately one-third of the simulation box volume, repurposing unused regions at lower redshifts and maximizing the utility of the simulation data \citep{Jain2000}. As shown in \cite{EC_higherorderWL_Ajani2023}, $85$ out of the $256$ light cones can be considered fully independent, each sharing less than $1\%$ of DM halos across different lines of sight at $z_s = 1$ (Sect. \ref{subsec:maps}).

Reusing the same simulation box for multiple light cones may slightly underrepresent rare events in the convergence tails, especially in high- or low-$\kappa$ observables, such as peak counts or voids \citep{Garrison2019, Rasera2022, Chen2024}. Although randomization reduces sample variance, residual correlations persist but are largely mitigated by galaxy shape noise (Sect. \ref{subsec:GSN}), which suppresses systematic variance differences \citep{J.Liu2015}.

\subsection{Convergence maps}\label{subsec:maps}
To accurately simulate the weak gravitational lensing signal, it is essential to trace the paths of light rays across the multiple lens planes constructed from our light cones. The mass maps produced in the first step i-\mapsim were used as input to perform multiplane ray-tracing calculations through ray-\mapsimm, as done in several studies such as \cite{Petri2016, Petri2017, Giocolietal2017, Giocolietal2018_mio, Hilbert2020}. In this subsequent phase, the routine constructed the lensing convergence map using the Born approximation by summing the surface mass density from each lens plane along the line of sight, weighted according to the lensing kernels $W(\chiD,\chiS)$. More details are given in Sect. \ref{subsec:wlform}.

In the Born approximation, deflections are integrated along a straight-line path rather than by iteratively displacing rays at each lens plane. This method keeps the rays naturally aligned with the simulation grid, avoiding the need to interpolate the projected matter density at the exact ray positions \citep{Hilbert2020}. Unlike the exact multiplane ray-tracing method, the Born approximation avoids solving the Poisson equation (Eq. \ref{eq:Poisson_equation}), making it computationally more efficient \citep{Giocolietal2016}. Studies such as \cite{Giocolietal2017} and \cite{Castroetal2018} confirm that this method provides an accurate estimation of the convergence power spectrum and the probability distribution function (PDF) down to sub-arcminute angular scales. Moreover, \citet{Schafer2012} showed through a perturbative expansion that the Born approximation remains highly reliable for WL analyses even at very small scales ($l \geq 10^4$) \citep{Giocolietal2018_mio}. Although post-Born corrections improve multiplane ray-tracing accuracy in such cases, their impact on void statistics is negligible \citep{Ferlito2024}.
\begin{figure}[ht]
\centering
\includegraphics[width=\columnwidth]{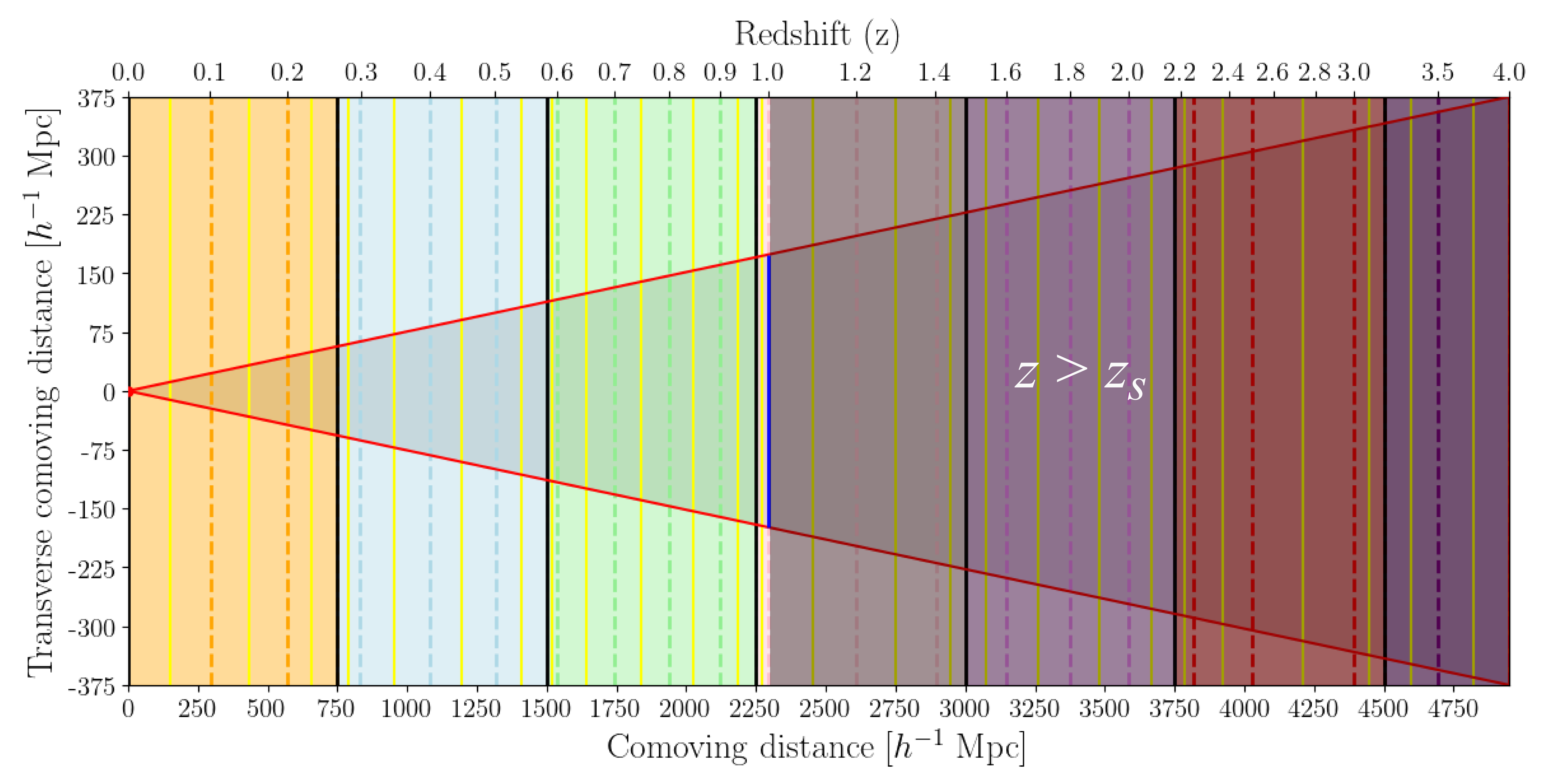}
\caption{\small{Example of light-cone construction with \mapsim from \( z=0 \) to \( z=4 \) for a \(\Lambda\)CDM simulation with a box size of \( 750 \, h^{-1} \, \mathrm{Mpc} \), tiled to encompass the \( 5 \times 5 \, \mathrm{deg}^2 \) light cone. The light cone geometry is depicted by the red solid line, expanding until its transverse comoving size matches that of the simulation box. Different colors represent different boxes, while the total of $21$ different snapshots are defined by the dashed lines. The source plane (blue solid line) for ray tracing is placed at \( z_s = 1 \). The snapshots within \( z_{\text{obs}} < z < z_s \) are the $13$ utilized ones, while the gray region indicates the snapshots with \( z > z_s \) that are not considered in this study. The yellow solid lines represent the $27$ lens planes used, onto which the particles are projected. Among these, only those within the considered snapshots are utilized for constructing the convergence maps used in this work, positioned from \(z=0.05\) to \(0.98\), respectively.}}
\label{fig:LC}
\end{figure}

In this framework, to compute the convergence map for each lens plane, we used the discretized form of Eq. \eqref{eq:kappa_born}\footnote{Note that the result is equivalent to the derivation made under the thin-lens approximation within the snapshot where the plane is constructed \citep{Jain2000}.}. Introducing the critical surface density in comoving units for a lens plane at redshift \(z_l\) and sources at \(z_s > z_l\), defined as
\begin{equation}\label{eq:sigma_crit_comoving}
\Sigma^{c}_{\rm{crit},l,s}(z_l, z_s) \equiv \frac{c^2}{4 \pi G} \frac{\fS a(z_l)}{\fD \fDS}=\frac{c^2}{4 \pi G} \frac{a(z_l)}{W(\chi_s, \chi_l)} \, .
\end{equation}
we omputed the convergence on the $l$-th lens plane as \footnote{In WL studies, projected densities and distances are typically expressed in comoving units. For instance, the critical surface mass density for lensing in comoving units, $\Sigma^{c}_{\rm{crit},l,s}(z_l, z_s)$, is related to its counterpart in physical units, $\Sigma^{ph}_{\rm{crit},l,s}(z_l, z_s)$, through the relation $\Sigma^{c}_{\rm{crit},l,s} = \Sigma^{ph}_{\rm{crit},l,s} (1 + z_l)^{-2}$ \citep{Umetsu20}.}
\begin{equation}\label{eq:K_sigma}
\kappa^l(\vtheta, \chiS) = \frac{\Sigma^l(\vtheta)}{\Sigma^{c}_{\rm{crit},l,s}} \, .
\end{equation}
At the end, the total convergence map is
\begin{equation}
\kappa(\vtheta, \chiS) = \sum_l \kappa^l(\vtheta, \chiS)=\sum_l \dfrac{\Sigma^l(\vtheta)}{\Sigma^{c}_{\rm{crit},l,s}} \, .
\end{equation}

This ray-tracing configuration precisely simulates the gravitational lensing effects caused by underdense regions in the Universe, facilitating an in-depth examination of lensing profiles from cosmic voids in the context of MG models. As shown in Fig. \ref{fig:LC}, in this work, the convergence maps were constructed from the corresponding light cones, emulating observational conditions by placing the background sources plane at \( z_s=1 \). This setup provides a continuous matter field over the specified FOV, segmented into $2048\times2048$ pixels, yielding a pixel resolution of approximately \( 9 \, \mathrm{arcsec} \). The origin of each map coordinate system (R.A., Dec.) is positioned at the center such that each side spans from $-150$ to $+150$ $\mathrm{arcmin}$. The redshift of the background source has been chosen because it is estimated to be the peak of the source probability distribution function, $n(z_s)$, for the \textit{Euclid} wide photometric survey \citep{Scaramella2022_Euclid_collab, EC_higherorderWL_Ajani2023, EC_Mellier_Overview2024}.

\subsection{Galaxy shape noise}\label{subsec:GSN}
Galaxy shape noise (GSN) represents a fundamental challenge in WL studies. It arises from the intrinsic random distribution in ellipticities and orientations of background galaxies, which heavily dominate the observed correlations in galaxy shapes caused by gravitational lensing \citep{Kilbinger2015ReviewCosmologyCosmicShearObservations}. GSN introduces a stochastic noise component that causes significant variability in the reconstruction of convergence maps, which must be addressed to accurately identify under- (valleys) and over-densities (peaks) regions in WL fields \citep{Lin&Kilbinger2015}. To simulate observational conditions, GSN is added to the simulated WL convergence maps as a Gaussian random field, specifically by superimposing random values drawn from a Gaussian distribution for each pixel \citep{VanWaerbeke2000}.

On each pixel of the map $\kappa(\vtheta)$, we added GSN, $n(\vtheta)$, modeled as a Gaussian random field with a top-hat filter with a size that corresponds to the pixel area $A_{\mathrm{pix}}=\theta^{2}_{\mathrm{pix}}$ in $\mathrm{arcmin}^{-2}$. According to \cite{Lin&Kilbinger2015} its variance is given by 
\begin{equation}
\sigma_{\text {pix }}^{2}=\frac{\sigma_{\epsilon}^{2}}{2} \frac{1}{n_{\mathrm{gal}} A_{\mathrm{pix}}} \, ,
\end{equation}
where $\sigma_{\epsilon}^{2}=\left\langle\epsilon_{1}^{2}\right\rangle+$ $\left\langle\epsilon_{2}^{2}\right\rangle$ is the variance of the intrinsic ellipticity distribution of the source galaxies and $n_{\mathrm{gal}}$ is the source galaxy number density. We assumed a \textit{Euclid}-like setup \citep{Scaramella2022_Euclid_collab, EC_higherorderWL_Ajani2023, EC_Mellier_Overview2024}, so $\sigma_{\epsilon}^{2}=0.3$ and $n_{\mathrm{gal}}=30 \ \mathrm{arcmin}^{-2}$ at $z_s=1$. In this way, we obtained a noised convergence map $\kappa_{n}(\vtheta)=\kappa(\vtheta)+n(\vtheta)$.

The inclusion of GSN significantly contaminates WL maps with noise \citep{Lin&Kilbinger2015}. This affects key statistics such as void abundance and shear profiles \citep{Davies2019SelfSimilpeaks, Davies2021OptimalVoidFinderinWeakLensing}. The GSN, however, can be reduced by applying a Gaussian smoothing filter characterized by a smoothing length $\theta_{\mathrm{G}}$ \citep{Kilbinger2015ReviewCosmologyCosmicShearObservations}. In this work, we used this approach to minimize GSN contamination in tunnel void statistics, ensuring consistency between the maps, with and without GSN.

Choosing an optimal value for $\theta_{\mathrm{G}}$ depends on the specific analysis, but it usually ranges between $1$ and $5$ arcmin. Previous studies have shown that cosmological constraints derived from WL peaks can be improved by employing multiple smoothing scales \citep{J.Liu2015}. However, it has also been demonstrated that a single intermediate smoothing scale can provide a balanced approach for specific WL peak analyses while remaining robust against GSN contamination \citep{Davies2019SelfSimilpeaks}. Selecting a small $\theta_{\mathrm{G}}$ retains information from smaller scales within the WL map but leaves a significant level of GSN contamination. In contrast, a large $\theta_{\mathrm{G}}$ suppresses noise more effectively but also erases small-scale features. Thus, selecting an optimal value requires balancing noise reduction with the preservation of meaningful WL signals.

In this work, we initially tested three smoothing scales: \( \theta_{\mathrm{G}} = 1 \), \( 2.5 \), and \( 5 \, \text{arcmin} \). To evaluate the robustness of void detection under different noise conditions, we considered each case with and without GSN. The results align with \cite{Davies2019CosmologicalTestGravityWeakLensing} and \cite{Davies2021OptimalVoidFinderinWeakLensing}, showing that small smoothing scales (\(\theta_{\mathrm{G}}=1\) arcmin) preserve fine structures but amplify noise, increasing variance in WL void statistics. Large scales (\(\theta_{\mathrm{G}}=5\) arcmin) reduce noise but erase meaningful features, lowering void detection sensitivity. Intermediate smoothing scales provide the best balance, minimizing noise while preserving statistical significance. Moreover, as demonstrated by \cite{Weiss2019}, such intermediate scales help to reduce the discrepancies between the employed simulations (which, in our case, do not include baryonic physics) and fully hydrodynamic simulations. For these reasons, we finally decided to carry on the analysis using \(\theta_{\mathrm{G}} = 2.5\) arcmin. Thus, we obtained the noised and smoothed convergence map as $\kappa_{2.5}(\boldsymbol{\theta}) \equiv\left(\kappa_{n} * U\right)(\boldsymbol{\theta})$ using
\begin{equation}
U(\boldsymbol{\theta})=\frac{1}{\pi \theta_{\mathrm{G}}} \exp \left(-\frac{\theta^{2}}{\theta_{\mathrm{G}}^{2}}\right) \, ,
\end{equation}
with $U(\boldsymbol{\theta})$ being the Gaussian window function with $\theta_{\mathrm{G}} = 2.5 \ \mathrm{arcmin}$. This choice is particularly suitable for analyzing void shear profiles while retaining sufficient small-scale information within WL maps.
\begin{figure}
\centering
\includegraphics[width=\columnwidth]{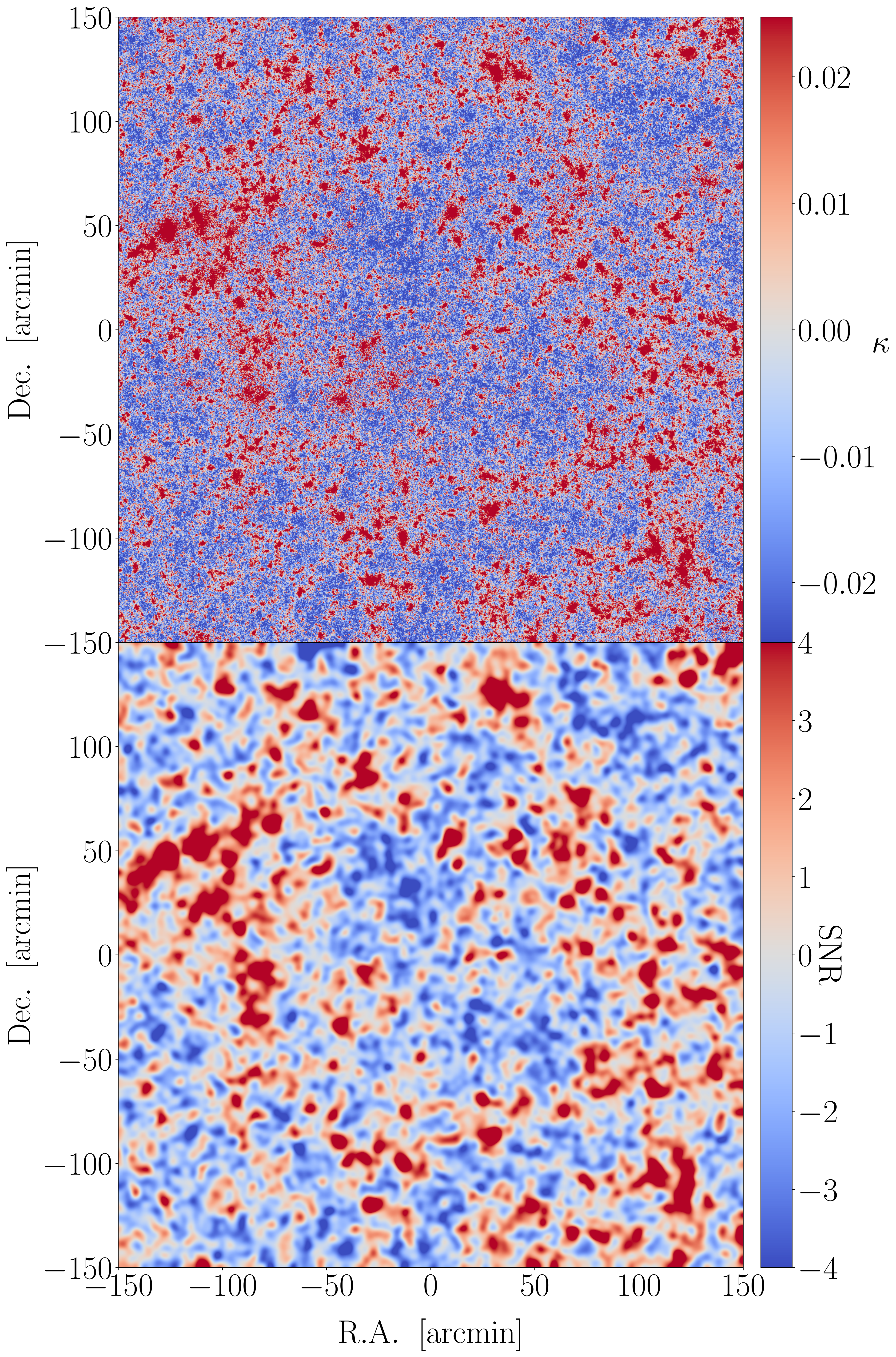}
\caption{\small{Example of a realization of the $\Lambda$CDM light cone, having a FOV of $5\times5$ deg$^2$ aperture and $z_s=1$. In the upper panel, we show the WL convergence map, and in the lower panel, the S/N of the same map with smoothing scale at $\theta_{\mathrm{G}}=2.5 \, \text{arcmin}$. The color scale in both panels represents the convergence value for each pixel in the top panel and the S/N value in the bottom panel, as indicated by the corresponding color bars.}}
\label{fig:K_SNR}
\end{figure}
We assumed that intrinsic ellipticities are uncorrelated between source galaxies. In this way, the noise after the smoothing can be also described as a Gaussian random field \citep{Bond1987}. According to \cite{VanWaerbeke2000}, its variance is related to the number of galaxies contained in the filter as
\begin{equation}
\sigma_{\text {noise }}^{2}=\frac{\sigma_{\epsilon}^{2}}{2} \frac{1}{2 \pi n_{\mathrm{gal}} \theta_{\mathrm{G}}^{2}} \, .
\end{equation}
This relation allows the derivation of the lensing S/N($\vtheta$)\footnote{The physical meaning of this ratio, sometimes also expressed as SNR, is that $S$ represents the number of photon counts on the sky area analyzed, while $N$ denotes the lensing background noise from sources in the background.} map of the noised and smoothed convergence map $\kappa_{2.5}(\vtheta)$ as
\begin{equation}
\text{S/N}(\vtheta) = \frac{\kappa_{2.5}(\vtheta)}{\sigma_{\text{noise}}(\theta_{\mathrm{G}})} \, ,
\end{equation}
where $\sigma_{\text{noise}}(\theta_{\mathrm{G}})$ is the standard deviation of the smoothed GSN map (without contributions from the WL convergence map, i.e., noise only). It varies depending on the smoothing scale used to identify the WL peaks or minima, whereas it is constant only if we assume that galaxies are uniformly distributed. In our framework, with $\theta_{\mathrm{G}}=2.5$ and $z_s=1$, $\sigma_{\text{noise}} \approx 6.18\cdot10^{-3}$.
\begin{figure}
\centering
\includegraphics[width=\columnwidth]{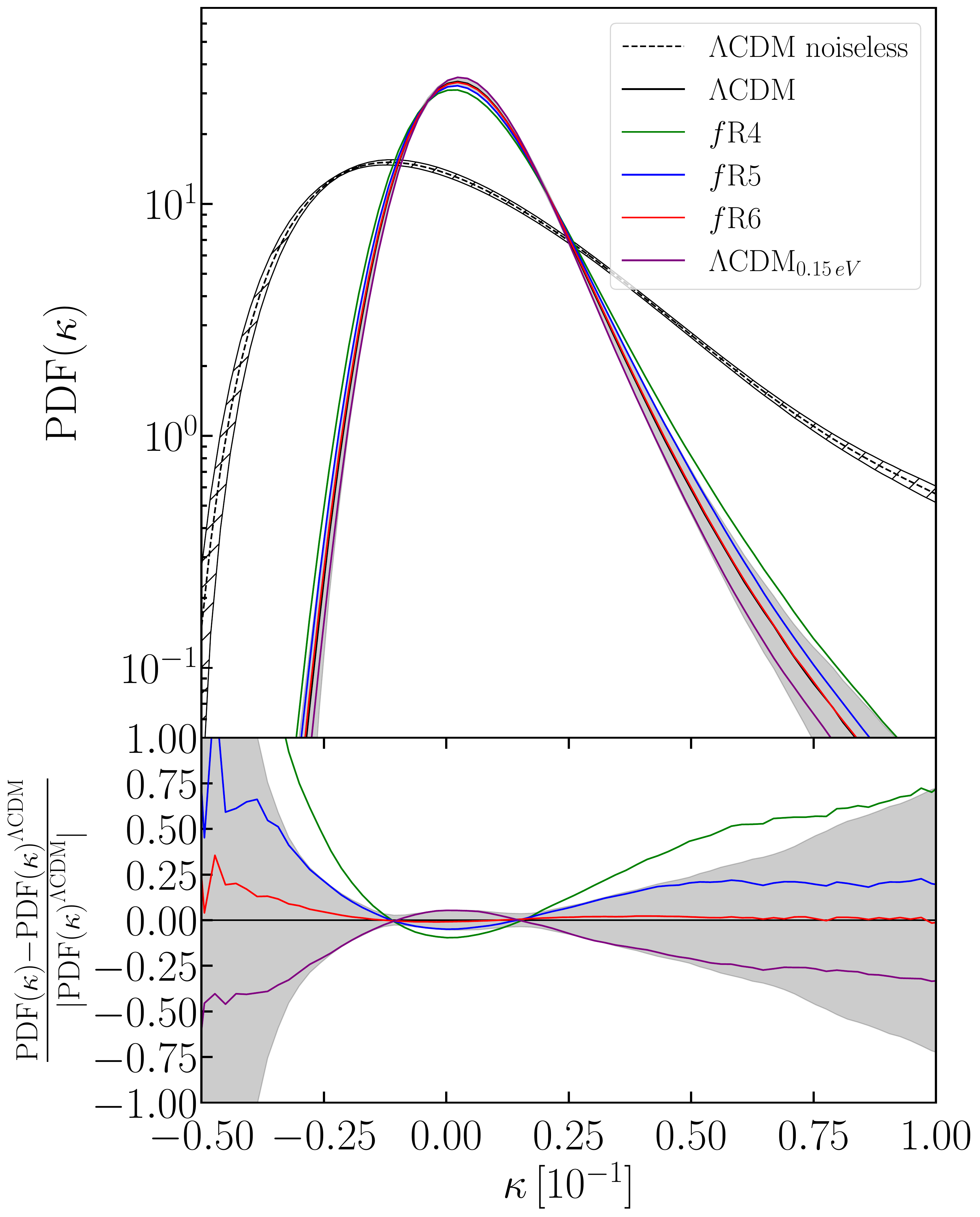}
\caption{\small{\textit{Top panel}: Average PDFs of the noised and smoothed convergence field from $256$ light-cone random realizations with redshift $z_s=1$, for all the analyzed models: $\Lambda$CDM (black), $fR4$ (green), $fR5$ (blue), $fR6$ (red) and $\Lambda$CDM$_{0.15 \, \mathrm{eV}}$ (purple). The black dashed line represents the average PDF without the addiction of GSN for the \lcdm model. \textit{Bottom panel}:  PDF noised and smoothed corresponding residuals computed with respect to the $\Lambda$CDM model. For the purposes of visualization, only the uncertainties of the noiseless and the reference $\Lambda$CDM measurements are reported in this plot as hatched gray and shaded gray areas, respectively.}}
\label{fig:pdf_k}
\end{figure}
Figure \ref{fig:K_SNR} illustrates the impact of applying the \textit{Euclid}-like GSN and a Gaussian smoothing scale of \( \theta_{\mathrm{G}}=2.5 \) arcmin to WL convergence maps. The top panel displays the noiseless convergence field for one of our $256$ light cones, while the bottom panel presents the corresponding S/N map. The smoothing enhances the detectability of large-scale structures by reducing small-scale noise while preserving the primary features of the field. As observed, the convergence field exhibits a complex distribution of over- and underdensities, while the S/N map highlights significant structures by suppressing uncorrelated noise. 

\subsection{Probability density function and power spectrum}\label{PDF_PK}
The statistical properties of the convergence field provide crucial insights into the underlying cosmological model. Therefore, we went on to analyze the one-point PDF and the angular power spectrum of the constructed maps.

The PDF of the convergence field quantifies the distribution of $\kappa$ values across the sky and is particularly sensitive to non-Gaussian features induced by structure formation. Figure \ref{fig:pdf_k} shows the mean PDF computed from $256$ light-cone realizations for all the analyzed cosmological scenarios, together with the mean PDF noiseless for the $\Lambda$CDM model. The noised and smoothed PDFs exhibit a subtle dependence on the cosmological model, while the difference is more pronounced when comparing the noised and smoothed $\Lambda$CDM PDF with its noiseless counterpart.

The bottom panel of Fig. \ref{fig:pdf_k} illustrates the residuals of the noisy and smoothed PDFs with respect to the standard $\Lambda$CDM model. In particular, to calculate the residuals of any quantity $X$ analyzed in this work, we adopted the following general formula
\begin{equation}\label{eq:residuals_general}
\Delta X = \frac{X^{\rm model}(r) - X^{\rm ref}(r)}{|X^{\rm ref}(r)|} \, ,
\end{equation}
where $X^{\rm model}(r)$ represents the profile or observable under investigation, while $X^{\rm ref}(r)$ corresponds to a fixed reference model for comparison. Here, the errors are computed as the standard deviation of the measurements across different realizations of the maps. Throughout the paper, the errors are properly propagated during the calculation of the residuals.

Looking at the residuals, we can appreciate how the deviations from the $\Lambda$CDM model become more relevant for $\kappa < 0$. Effectively, the largest deviations occur in the negative $\kappa$ range, where $f(R)$ models display an excess probability compared to $\Lambda$CDM, while the massive neutrino model exhibits a suppression. The lower PDF values in underdense regions strengthen fractional deviations. MG models amplify structure formation, deepening underdensities and enhancing non-Gaussian features. Massive neutrinos, conversely, suppress structure growth, shifting the PDF toward lower contrast. Since these effects are more pronounced in the negative $\kappa$ tail, we expect WL from tunnel voids to represent a key tool for testing alternative cosmologies as they can capture the differential impact of MG and neutrinos.

\begin{figure}
\centering
\includegraphics[width=\columnwidth]{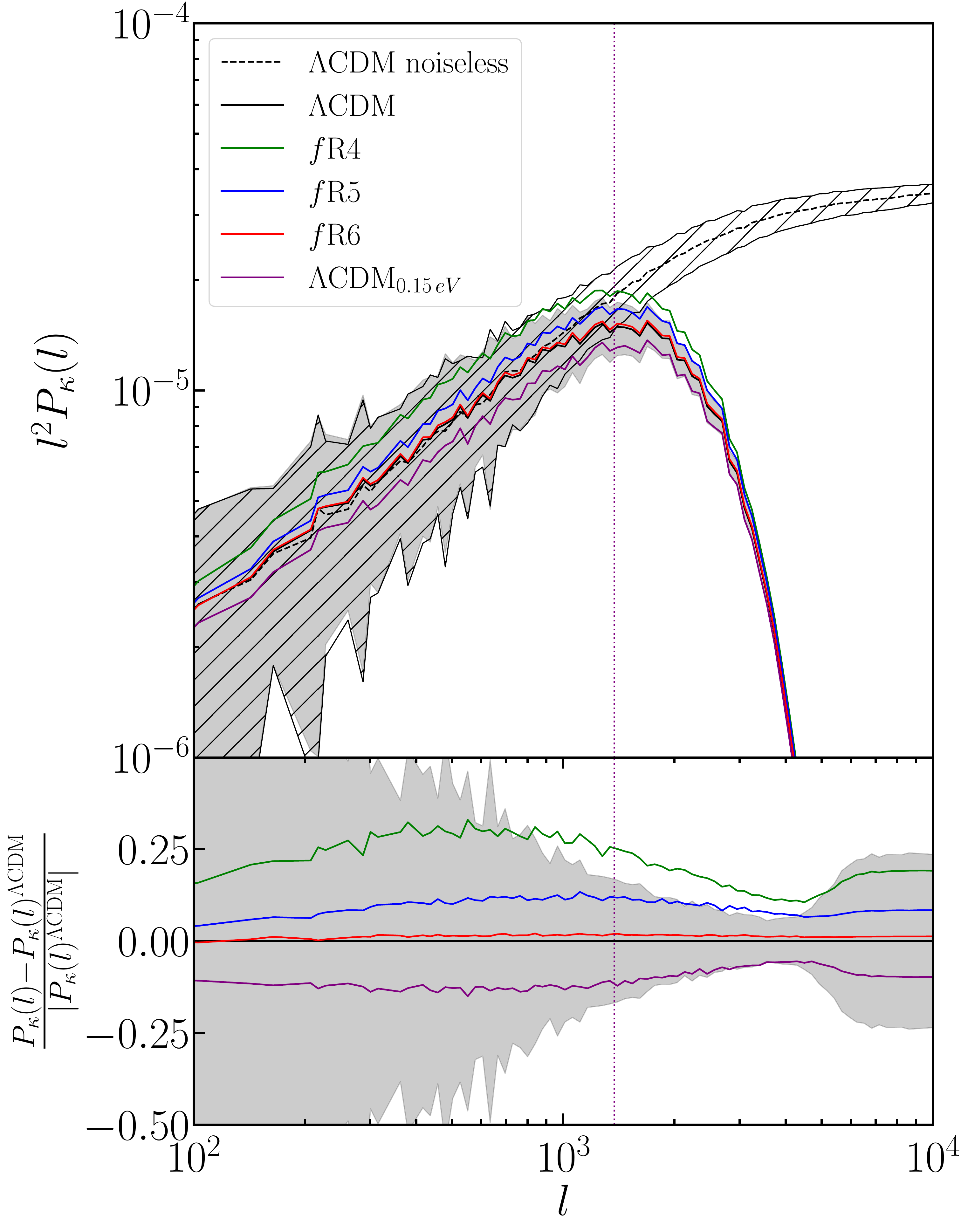}
\caption{\small{Average angular power spectra of the noised and smoothed convergence field from $256$ light-cone random realizations with redshift $z_s=1$, for all the analyzed models (top panel) and the correspondent residuals with respect to the reference $\Lambda$CDM model (bottom panel). Line styles and color schemes are the same as used in Fig. \ref{fig:pdf_k}. The dotted vertical line marks the angular mode of the characteristic smoothing scale in the Fourier transform of the $\kappa$ field.}}
\label{fig:p_k}
\end{figure}

The angular power spectrum of the convergence field, $P_{\kappa}(l)$, provides a complementary statistical measure by characterizing the variance of the $\kappa$ maps fluctuations, describing how power is distributed across different angular scales. In Fig. \ref{fig:p_k}, it is shown as a function of the multipole moment $l$, which corresponds to the inverse angular scale on the sky normalized for the field size ($2\pi/\theta_{FOV}$). It is computed as the Fourier transform of the input map squared and binned in multipole space.

In particular, Fig. \ref{fig:p_k} shows the mean power spectrum computed from $256$ convergence maps for all the analyzed cosmological scenarios alongside the noiseless mean power spectrum for the $\Lambda$CDM model. The difference between the models can be better appreciated in the bottom panel, where the relative differences of the power spectra with respect to $\Lambda$CDM are presented. The largest deviations from the \(\Lambda\)CDM model occur at the intermediate scales, in the multipole range \( [300 - 1000] \). This range corresponds to angular scales between $[11.46, 3.44]$ arcmin. The impact of MG and massive neutrinos is maximized in this regime. \( f(R) \) models enhance structure formation, leading to an increase in power on these scales. Massive neutrinos suppress clustering, causing a damping of power. We can notice how these competing effects lead to significant deviations in the convergence power spectrum.

At larger scales (\( l \lesssim 300 \)), the Universe remains in the linear regime, where deviations are smaller due to the weaker impact of nonlinear modifications. At smaller scales (\( l \gtrsim 1000 \)), nonlinear effects dominate, and Gaussian smoothing starts to suppress the structure-dependent variance.

In Fig. \ref{fig:p_k}, the vertical line marks the characteristic smoothing scale in Fourier space (\( l = 1375.10 \)). Beyond this limit (\( \theta_G = 2.5 \) arcmin), the signal is attenuated and significantly deviates from the noiseless power spectrum, reducing sensitivity to cosmological model differences. This behavior motivates us to investigate in depth the effect of WL on larger scales and, given the findings from the analysis of the convergence PDF and power spectrum, to focus especially on voids. 

\section{2D tunnel void finder}\label{sec:void_finder}
\begin{figure*}
\centering
\includegraphics[width=1.0\linewidth]{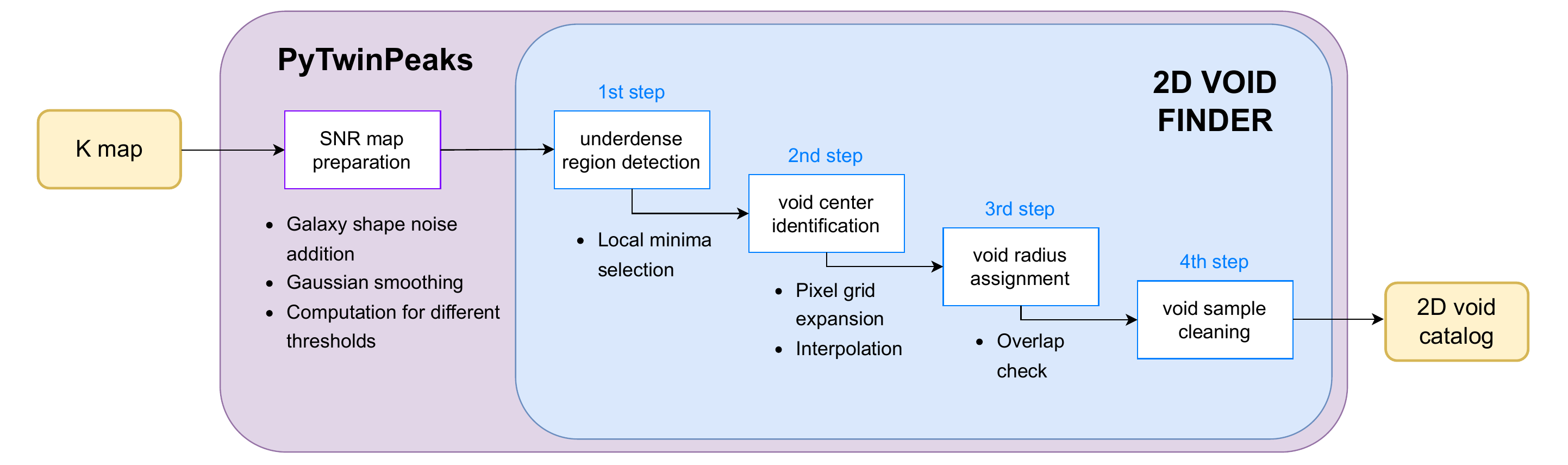}
\caption{\small{Diagram showing the four main steps of the 2D void finder algorithm that allows the identification of voids from a WL convergence map.}}
\label{fig:finder_scheme}
\end{figure*}

One of the main challenges in the study of cosmic voids lies in their definition and, consequently, in their identification. Cosmic voids are not luminous structures; instead, they can be defined only by the distribution of matter tracers, such as galaxies, which predominantly reside along their boundaries. This necessitates the use of ad-hoc techniques to reconstruct void shapes and centers from the tracers, called void finder algorithms. The absence of a universally accepted, first-principles definition of voids makes their identification via void finders highly dependent on the specific goals and type of analysis being performed \citep{Colberg2008, Cautun2018, Davies2019CosmologicalTestGravityWeakLensing, Davies2021OptimalVoidFinderinWeakLensing}.\\
\indent In this work, we developed a novel 2D void finder algorithm for WL analyses, guided by these insights, incorporating features from existing methods in the literature while addressing their limitations to enhance both stability and effectiveness. Specifically, we implement a new tunnel finder based on WL absolute minima, adding GSN to our maps. We have made it publicly available (see Sect. ‘Data availability’). The structure and the main steps of the code are schematically represented in Fig. \ref{fig:finder_scheme} and analyzed in detail in the following.

\subsection{\texttt{PyTwinPeaks}: Reconstruction of connected regions}\label{subsec:conn_regions}

The processed convergence maps serve as the input for our 2D void finder algorithm, which detects underdense regions based on their S/N. The algorithm called \texttt{PyTwinPeaks} is a Python version of \texttt{TwinPeaks} c++-based code \citep{Giocoli2018_twinpeaks}, a peaks/valleys finder that analyzes overdense and underdense S/N regions. In this work, we used \texttt{PyTwinPeaks} to reconstruct underdense lines-of-sight pixel regions at different convergence S/N thresholds, improving its efficiency. This algorithm was later expanded and transformed into a dedicated WL tunnel void finder.

The first phase of the algorithm consists of a reconstruction of the valleys: it begins with the creation of a mask that selects pixels in the S/N map below the chosen threshold. Next, the reconstruction is done via an affiliated component analysis, identifying pixels belonging to the same connected region. The algorithm builds up a total map as the sum of layers, where each layer represents a connected region with a S/N value that is below the selected threshold.

From each connected region, the code extracts $16$ topological features, including perimeter, area, and eccentricity. It also records the $(x,y)$ coordinates of the weighted centroid, computed using the S/N values of individual pixels as weights, and the coordinates of the local minimum, defined as the pixel with the lowest S/N value. The coordinates, originally in pixels, are converted to arcminutes using the relation
\begin{equation}\label{eq:conversion_pix}
x_\mathrm{arcmin} = \left(x_\mathrm{pix} - \frac{n_\mathrm{pix}}{2}\right) \cdot l_\mathrm{arcmin},
\end{equation}
where $x_\mathrm{pix}$ represents the coordinate in the S/N map in pixel units, $n_\mathrm{pix}$ is the number of pixels per side of the map, and $l_\mathrm{arcmin} = \mathrm{FOV}_\mathrm{arcmin} / n_\mathrm{pix}$ denotes the pixel size in arcminutes.

Once the underdense regions are reconstructed based on a specified threshold, overlapping cases between nearby regions are handled. Each connected region is initially approximated as a circle with a preliminary radius $R_{\mathrm{pr}}$, which corresponds to the equivalent radius of the circle with the same area. Although the regions are disjoint by construction, their circular representation may lead to overlaps. Two regions merge into a larger structure if the distance between them satisfies the condition
\begin{equation}
D_\mathrm{pr} < \frac{R_{\mathrm{pr},1} + R_{\mathrm{pr},2}}{2}.
\end{equation}

The algorithm repeats this procedure over a given threshold range, generating a catalog of underdense regions. The number of thresholds tested is adjustable and affects both the accuracy of the search and the computation time. To identify the connected pixel regions resulting from the projection of underdense lines-of-sight, we analyzed $11$ threshold levels from $\text{S/N} = -5$ to $\text{S/N} = 0$ with a step of $0.5$. The resulting catalog contains the properties of each connected region for each threshold. 

\subsection{Void center assignment}\label{subsec:centers}
This step is crucial, as it determines the final catalog of void centers in the projected S/N field and significantly affects the resulting statistics. As mentioned earlier, at the end of the first phase, the catalog includes the positions of the local minima, the weighted centroids, and the areas of each connected region across different thresholds. Each threshold produces a different selection of voids, and a center can be assigned to each connected region identified at a given threshold.

\begin{figure*}
\centering
\includegraphics[width=1.0\linewidth]{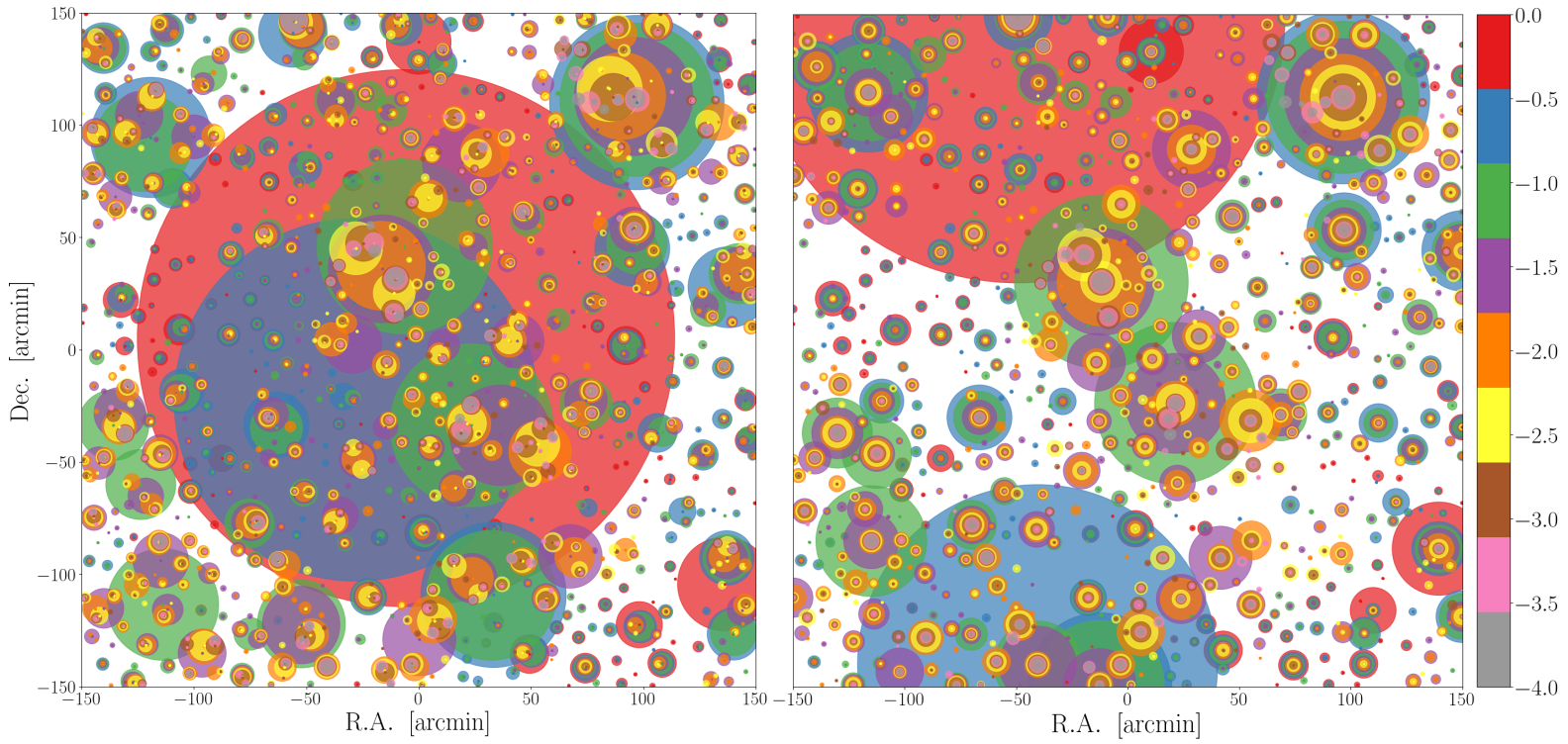}
\caption{\small{\textit{Left panel}: Graphical representation of all 2D voids with their preliminary radii of each threshold produced by the same map shown in Fig. \ref{fig:K_SNR}, centered at the S/N-weighted centroids of the connected regions. \textit{Right panel}: Same as the left panel, but with voids centered at the local minima of the connected regions. In both, each void of a given threshold is represented with the colored area of the same color, this color varies depending on the S/N threshold as shown in the colorbar.}}
\label{fig:centers}
\end{figure*}

Among the many void finders in the literature, the two primary approaches for assigning the void center (including the ones explored in this work) are through the centroid \citep[see][]{Padilla2005, Platen2007, Neyrinck2008, Sutter2015} and the local minimum \citep[see][]{SanchezDES2017, Nadathur2019_REVOLVER, Davies2021OptimalVoidFinderinWeakLensing, HangDESI2021_voidsISW, KovacsDES2022} of the connected region. In Fig.~\ref{fig:centers}, we illustrate the hierarchical structure of voids identified across all thresholds, using underdense regions centered on weighted centroids (left panel) and local minima (right panel), colored accordingly for different S/N values. The void sizes shown here are not the final results of our finder; their dimensions are determined by the area of the region at the highest threshold, which does not fully capture the complex morphology of a tunnel void.\\
\indent Each approach has its strengths and limitations, and the choice between them depends on the specific use case and scientific objective. For example, although the centroid method weights the S/N values, the assigned center may occasionally fall outside the connected region. At higher (still negative) S/N values, connected regions can have elongated or irregular shapes, and in extreme cases (e.g., a crescent-shaped underdensity), the centroid may fall within or too close to a positive S/N (overdense) region. This misplacement contaminates the tangential shear profile and distorts the extracted signal. Moreover, as shown in Fig.~\ref{fig:centers}, the centroid method introduces a continuous shift of void centers depending on the selected S/N threshold, which affects the total number of 2D voids. In fact, the positions of void centers directly influence how underdense regions merge. On the other hand, centroids help reduce the impact of GSN at lower S/N values. Using instead local minima as void centers maximizes the tangential shear signal and resolves the topological issues associated with centroids. However, voids centered on local minima are more susceptible to noise fluctuations \citep{Davies2021OptimalVoidFinderinWeakLensing}.\\
\indent In light of these considerations, in the new version of \texttt{PyTwinPeaks}, we developed a method for void center assignment based on what we call absolute minima. This strategy is inspired by a common technique in 3D void finders known as the watershed algorithm \citep[see][]{Platen2007, N&H2015WatershedVoidFinders-I}, here adapted for 2D voids. The algorithm tracks local minima across multiple thresholds, starting from the lowest and checking whether each minimum persists (i.e., whether it falls within the area outlined by the next threshold) as the threshold increases up to a user-defined limit. A minimum that persists across all levels is classified as an absolute minimum.\\
\indent The initial threshold sets the depth at which void centers are identified and directly impacts the depth of the final voids. By construction, this depth affects the amplitude of the tangential shear signal extracted from those regions. A balanced choice ensures a statistically significant number of 2D voids while preserving the strength of the WL signal. The final threshold is equally important. A low value causes absolute minima to coincide with local minima, creating small and fragmented voids. Conversely, a high value enlarges underdense regions along the line of sight, leading to the excessive merging of multiple structures. The final threshold must be carefully chosen to ensure reliable void statistics while avoiding overmerging. Contrary to the example shown in Fig.~\ref{fig:centers}, after extensive testing, we found that the range $\text{S/N}=[-4, -3]$ is optimal for setting the minimum and maximum selection thresholds: this technique produces tangential shear profiles with the deepest signals and the smallest uncertainties. The latter is computed from the scatter among different void profiles in the catalog, as we discuss in Sect.\ref{sec:statistics}.\\
\indent For the analysis presented in this work, the implementation of this method proved effective in resolving the limitations of the two previously mentioned approaches. Absolute minima in the S/N field define the void centers, ensuring an optimal shear signal. GSN contamination is simultaneously mitigated, together with map smoothing, by analyzing underdensities across multiple thresholds and checking for center persistence in higher S/N regions. The final catalog generated by this phase consists of underdense regions centered on a persistent minimum identified at the lowest threshold, with sizes determined by the area of the region at the highest threshold.

\subsection{Radius assignment and cleaning}\label{subsec:radii}
The next phase of the developed void finder algorithm assigns the final radius to the void centers identified in previous steps. To adapt this radius to the statistical signal of underdense lines-of-sight, such as tunnel voids, we impose each void to correspond to a circle enclosing the exact S/N final value. At the same time, the radius must be adjusted to account for the spatial resolution and boundary constraints of the convergence map.

The algorithm initializes a grid of $3\times3$ pixels around each void center, whose coordinates are converted from arcminutes to pixels using the inverse formula of Eq. \eqref{eq:conversion_pix}. A circular mask\footnote{The choice of a circular mask is motivated by the initial definition of tunnel voids, which are approximated as cylinders with equal circular bases, appearing as circles in 2D.} is applied to this grid, selecting all pixels enclosed by or intersected by the mask. The mask radius is set according to the relationship
\begin{equation}\label{eq:grid}
r_\mathrm{mask} = \frac{l_\mathrm{grid}}{2} - \frac{l_\mathrm{pix}}{2},
\end{equation}
where $l_\mathrm{grid}$ represents the grid size and $l_\mathrm{pix}$ denotes the pixel size, both measured in pixel units. The mask ensures high precision and computational efficiency by restricting the selection of relevant pixels.
\begin{figure}
\centering
\includegraphics[width=\columnwidth]{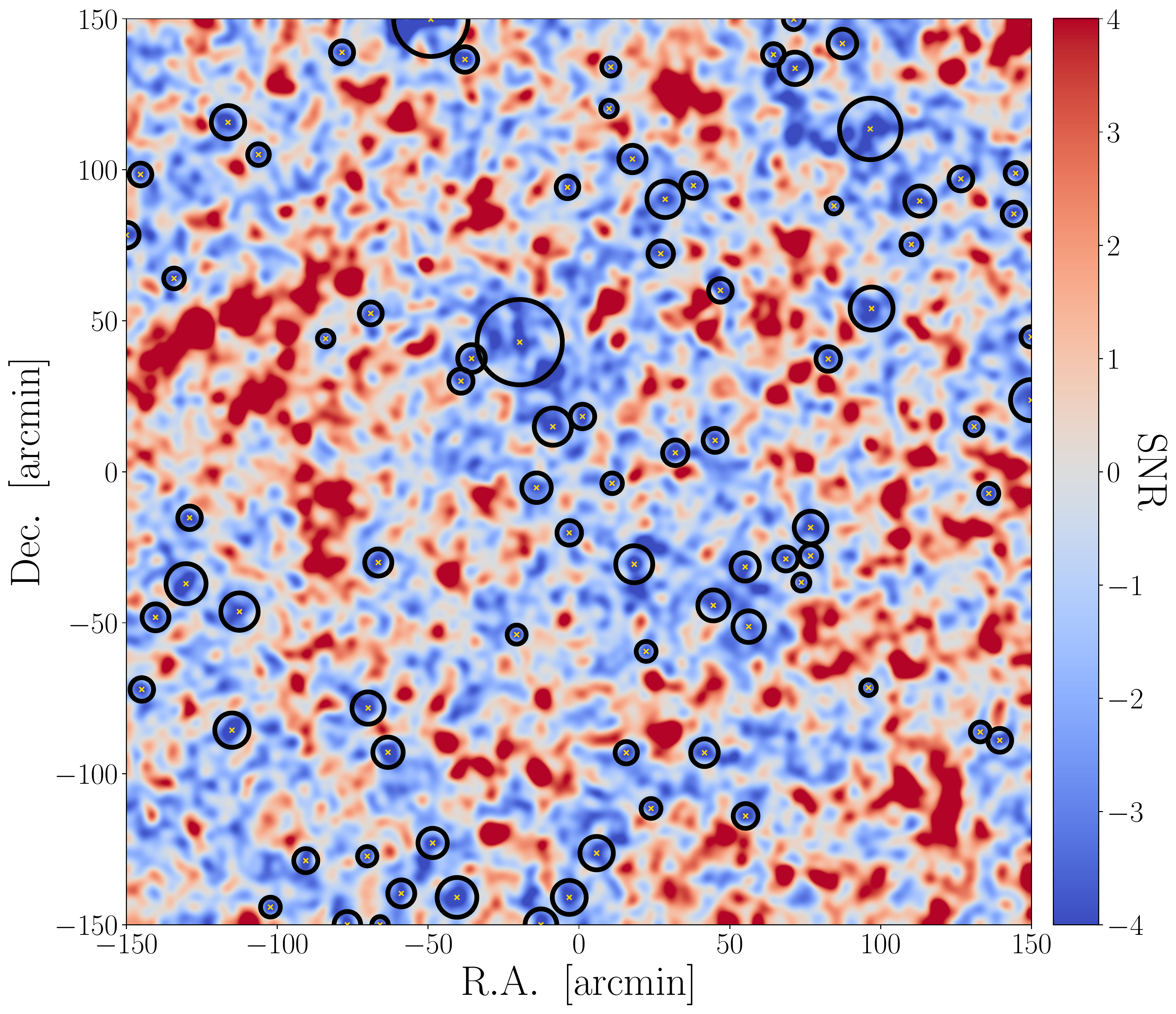}
\caption{\small{Example of the final tunnels catalog associated with the same S/N map of Fig. \ref{fig:K_SNR} and Fig. \ref{fig:centers}. Colors from blue to red represent areas with increasing S/N, as indicated by the colorbar on the right. The 2D voids identified by our algorithm are shown as areas delimited by black circles. The corresponding centers (absolute minima) are marked with yellow crosses.}}
\label{fig:finder}
\end{figure}

The algorithm expands the grid by adding a row of pixels on each side until the average S/N within the circular mask exceeds the selected final threshold. The final radius is determined by interpolating between the last two values, refining the estimate of the radius that precisely corresponds to the considered threshold. As the final threshold for the expansion of the pixel grid in the ray assignment, we used the value $\langle \text{S/N} \rangle = -2.5$ (see Sect. \ref{subsec:WL}).\\
\indent For voids located near the boundaries of the S/N map, removing them from the catalog would reduce statistical significance. To address this, the algorithm masks grid cells that extend beyond the boundaries of the map. Pixels outside the map are assigned a "not a number" (NaN) value, ensuring they do not influence the averaging process. This approach effectively extends the map while preserving the integrity of the void identification procedure.\\
\\
\indent We were then able to refine the void catalog by removing overlapping voids. With the final radius assigned to each underdensity, the last requirement is to eliminate voids that are not independent, as WL signals extracted from overlapping voids cannot be treated separately. Since an initial overlap check was performed earlier, this refinement is expected to have a limited impact on the final catalog.\\
\indent The identified voids are sorted in descending order according to their size. The algorithm checks whether the distance between the centers is less than or equal to $75 \%$ of the sum of the two radii: $D_\mathrm{centers} \leq 75\% \cdot (r_{p,1} + r_{p,2})$, where $r_{p,1}$ and $r_{p,2}$ represent the projected radius of the two voids. In this way, we can also correct for the void-in-void scenario, which happens when a void is completely contained in a larger one. The updated void list, ordered by size, prevents unnecessary removal of underdensities. In this way, when multiple voids overlap in sequence, only the central one is removed, preserving the largest and smallest, which remain non-overlapping.\\
\indent The final output of the algorithm is an ASCII file containing three columns. Each row stores the void center positions in the X and Y directions and the void radius, all measured in arcminutes.\\
\indent Figure \ref{fig:finder} shows, as an example, a S/N map from one of the $256$ convergence maps produced for the $\Lambda$CDM scenario, where circles mark the positions of the 2D voids. We can note that these voids correspond very well with the negative S/N regions. All negative regions on the map that are not associated with 2D voids are either considered part of a larger, spatially nearby void or are considered to be caused by S/N fluctuations. We underline how the chosen criterion for assigning the void radius is particularly effective in identifying voids as entirely negative regions surrounded by positive areas; namely, the void sizes result correctly expanded until they reach a WL positive signal. Additionally, this method optimizes the total computational time required. The void finder takes about 1 minute to produce the final void catalog from a single map and around 4 hours to analyze all $256$ maps, running on a local machine with $12$ cores, a maximum frequency of $4.7$ GHz, and a CPU scaling\footnote{Percentage of the maximum frequency of the processor currently being utilized to optimize power efficiency.} at $13\%$.

We also emphasize that the tunnel voids identified by this new 2D void finder result from a specific combination of choices. These include the selection of the GSN, the smoothing scale, the threshold range for the watershed method to distinguish absolute minima, and the final threshold for the radius assignment. Despite the presence of free parameters in the algorithms, which makes the final catalog susceptible to user choices, this flexibility makes the algorithm particularly suitable for application to real data from ongoing and future surveys.
 
\section{Tunnels void statistics}\label{sec:statistics}
\subsection{2D void size function}\label{subsec:VSF}
Compared to 3D voids identified in galaxy surveys, generally covering scales of tens of megaparsecs \citep[see, e.g.,][]{Contarini2022BOSS_vsf}, WL tunnel voids extracted in this work from S/N maps are characterized by relatively small sizes. Their angular extension in the sky-plane spans from $2$ to $15 \ \mathrm{arcmin}$.

The reduction in void size is expected due to the nature of WL tunnel voids. These voids form when photons from distant sources primarily pass through underdense regions, particularly at redshifts near the peak of the lensing kernel function. The observed underdensities result from multiple 3D voids that partially align with the observer's line of sight. The projection onto the sky plane does not preserve the original size of 3D voids. Instead, it highlights only the areas where the signal is dominated by negative convergence.

\begin{figure}
\centering
\includegraphics[width=\columnwidth]{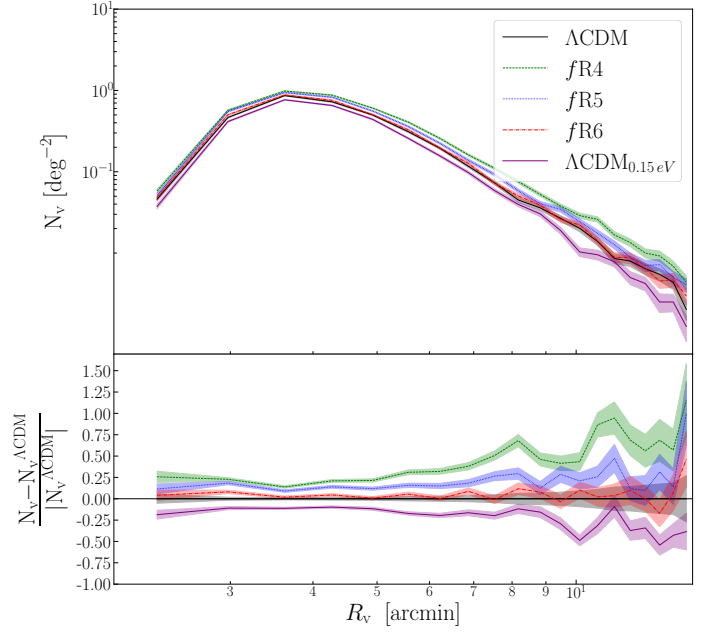}
\caption{\small{\textit{Top panel}: Average void size functions, per degree squared, of 2D WL voids from $256$ light-cone random realizations with redshift $z_s=1$, for the different analyzed models: $\Lambda$CDM (black), $fR4$ (green), $fR5$ (blue), $fR6$ (red), and $\Lambda$CDM$_{0.15 \, \mathrm{eV}}$ (purple). \textit{Bottom panel}: Corresponding residuals computed with respect to the $\Lambda$CDM model. In both panels, the shaded regions of the same colors around the curves indicate errors, computed as the standard deviation of counts in each bin divided by the number of maps.}}
\label{fig:vsf}
\end{figure}

In Fig. \ref{fig:vsf} we show the tunnel void size function (VSF), that is, the total number counts of 2D voids as a function of their projected radius $R_{\rm v}$ in arcminutes, averaged from the $256$ S/N map realizations, for each cosmological model. We restrict our analysis to the range $[2, 15] \ \mathrm{arcmin}$ to avoid the statistically rarest objects, which are characterized by large uncertainties. The error associated with void number counts is estimated as the scatter across independent realizations, normalized by the square root of the number of maps. On the top panel, we present the five VSFs, and, in the bottom panel, the corresponding residuals with respect to the $\Lambda$CDM model, computed as in Eq. \eqref{eq:residuals_general}.

The shape of all the VSFs is consistent with the one measured for 3D voids \citep{Hamaus2016voidsStatistics, Ronconi2017, Contarini2022Euclid, Contarini2022BOSS_vsf}, namely, it shows how small voids are more abundant with respect to the larger ones. The void radius, defined on the basis of the S/N in the 2D convergence map, reflects changes in the void interiors. Specifically, MG leads to larger, deeper voids, while massive neutrinos produce smaller, shallower ones \citep{Contarini2021_voids_MG_neutrinos}. This shifts the VSF accordingly, especially at large radii. At the small end, however, the peak remains fixed due to resolution limits: pixel size and smoothing set a lower bound on the detectable void size. A minimum radius cut and increased void mergers at large radii further impact the total tunnel void counts, leading to significant cosmology-dependent variations in VSF amplitude.

By analyzing different cosmological models, we note that all the size distributions are characterized by a similar shape and the main difference is in their amplitude. The observed trend for different cosmological scenarios aligns with expectations: in MG models, higher values of the parameter $|f_{R0}|$ lead to a greater increase in void counts. In the presence of massive neutrinos, the abundance of voids is suppressed. This happens because a stronger scalar field action, associated with higher $|f_{R0}|$ values, enhances the evolution of LSS, including cosmic voids. Conversely, larger neutrino masses suppress the growth of cosmic structures, reducing void formation. Consequently, the models $fR6$ and $\Lambda$CDM$_{0.15 \, \mathrm{eV}}$ are expected to be the most similar to the $\Lambda$CDM case, although they exhibit opposite trends. Specifically, up to $R_{\rm v} \sim 10\ \mathrm{arcmin}$, we observe an amplitude difference of $\leq 10\%$ for the $fR6$ model, while for $\Lambda$CDM$_{0.15 \, \mathrm{eV}}$, the difference is approximately $25\%$.

\subsection{Stacked tangential shear profile}\label{subsec:WL}
\begin{figure}
\centering
\includegraphics[width=\columnwidth]{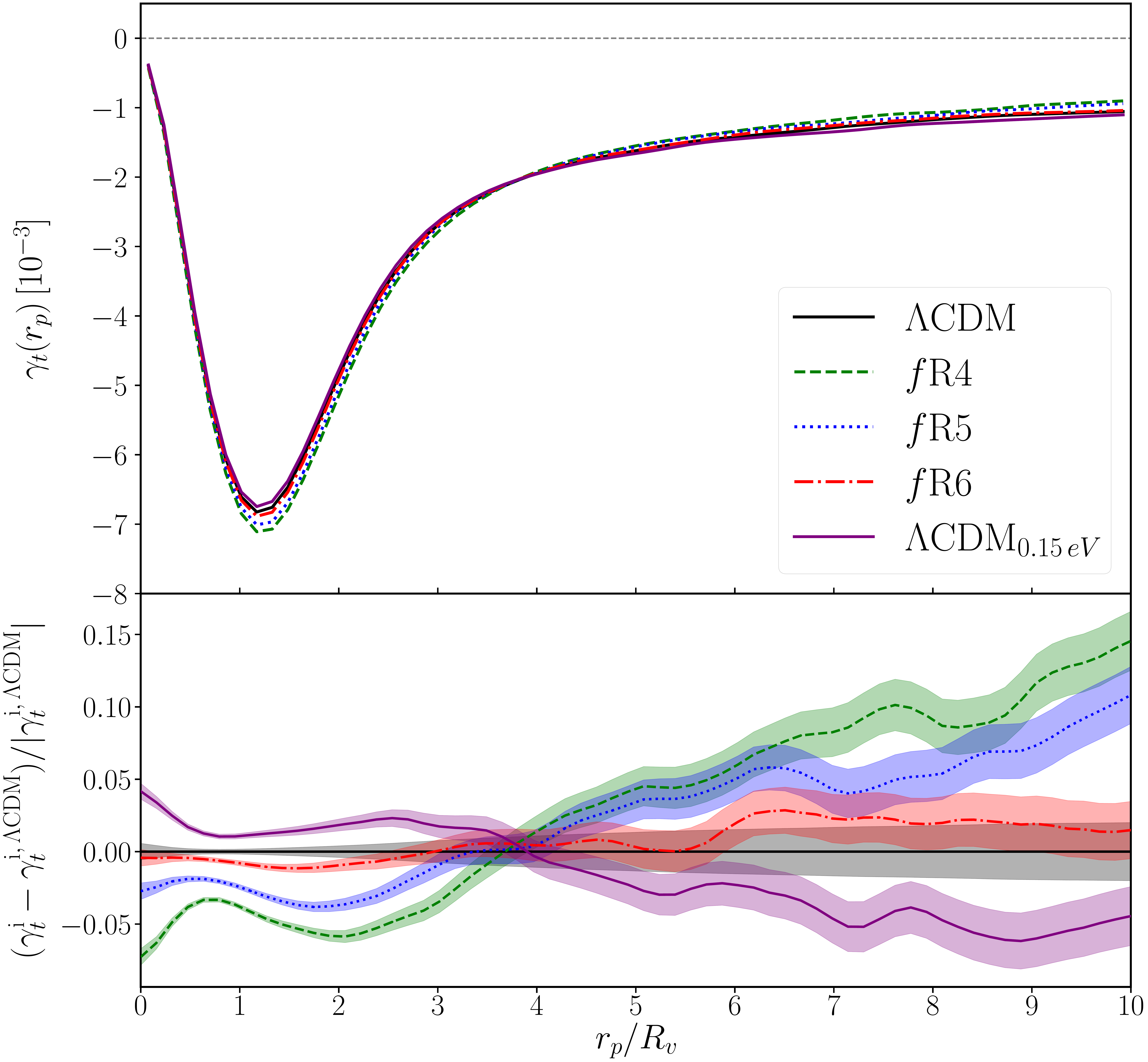}
\caption{\small{\textit{Top panel}: Averaged tangential shear profiles extracted from $256$ light-cone random realizations with redshift $z_s=1$, for the different analyzed models. \textit{Bottom panel}: Corresponding residual with respect to the profile measured in the $\Lambda$CDM scenario. Both the plots utilize the following color scheme: $\Lambda$CDM (black), $fR4$ (green), $fR5$ (blue), $fR6$ (red), and $\Lambda$CDM$_{0.15 \, eV}$ (purple). The shaded regions with the same colors around the curves represent the associated uncertainties, computed via the corrected covariance matrix (see Eqs. \ref{eq:gt_errors} and \ref{eq:alpha_Hartlap}). These are not displayed in the top panel for visualization reasons.}}
\label{fig:shear_1bin}
\end{figure}

The tangential shear profile of each WL tunnel void is extracted by expanding $64$ concentric circular shells around each void center in the noised and smoothed convergence map $\kappa_{2.5}(\vtheta)$, spanning from $r_{\rm min}=0$ to $r_{\rm max}=10 R_\mathrm{v}$. The tangential shear $\gamma_t(r_p)$ is computed for each radial bin using Eq.\ \eqref{eq:gamma_t}. Finally, we constructed the stacked tangential shear profile by averaging the profiles of different voids in a chosen sample.

As done in \cite{Davies2019CosmologicalTestGravityWeakLensing} and \cite{Davies2021OptimalVoidFinderinWeakLensing}, the uncertainty associated with the mean tangential shear profile of $N$ void profiles in the \( i \)-th bin is evaluated as
\begin{equation}\label{eq:gt_errors}
\sigma_i = \sqrt{\frac{C_{ii}}{N}} \, ,
\end{equation}
where \( C_{ii} \) is the \( i \)-th diagonal element of the covariance matrix calculated using the tangential shear profiles extracted from all the 256 S/N maps. We refer to Appendix \ref{app:correlation} for an in-depth analysis of the obtained covariance matrix. 

\begin{figure}
\centering
\includegraphics[width=\columnwidth]{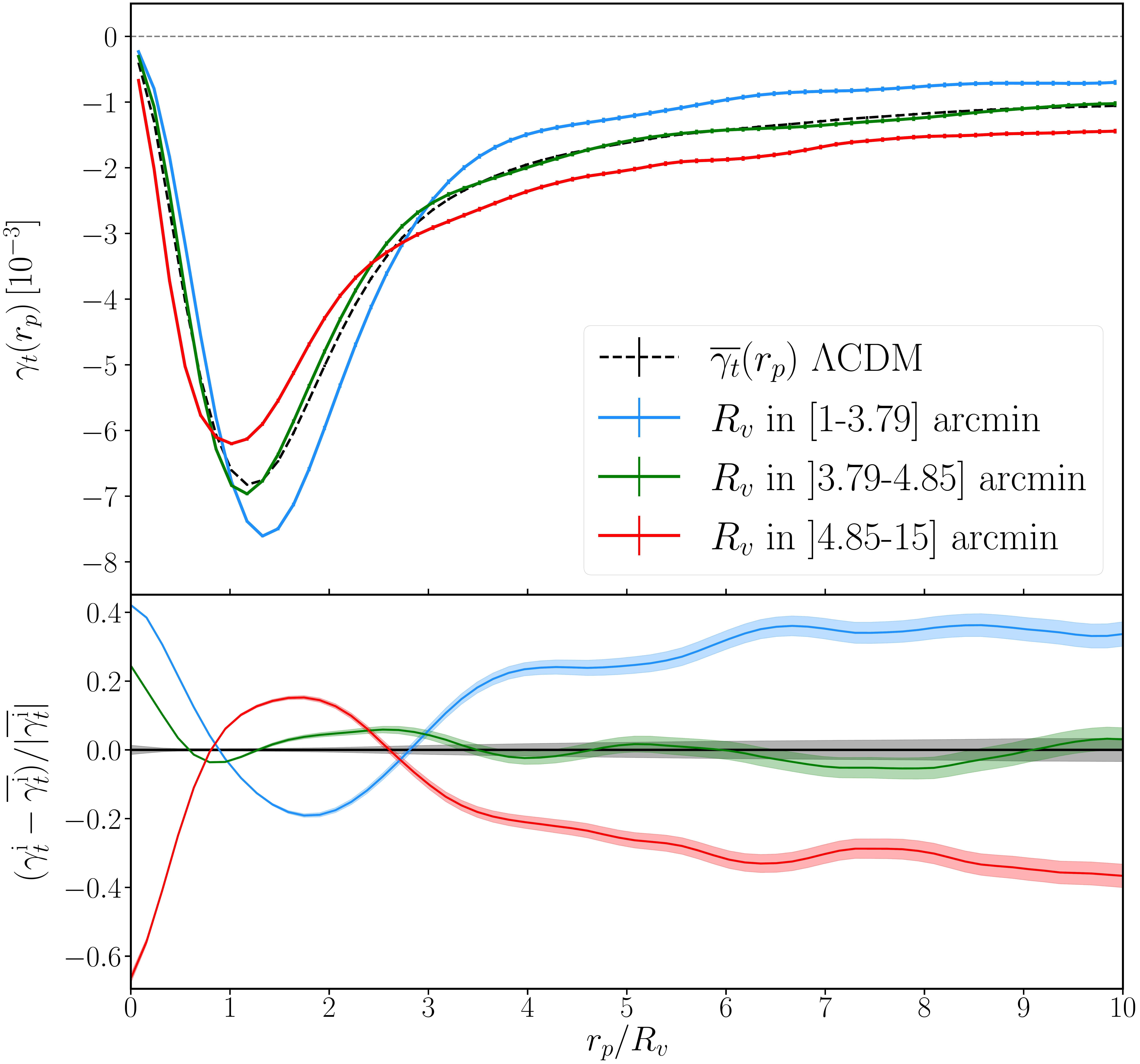}
\caption{\small{\textit{Top panel}: Averaged tangential shear profiles measured from samples of voids with different sizes in the standard $\Lambda$CDM cosmology: small (in light-blue), medium (green) and large (red). We report with a black dashed line the one computed using the whole void sample. \textit{Bottom panel}: Residuals between the profiles of each radius selection and the stacked void profile relative to the full sample. The shaded regions with same colors around the curves represent the uncertainty associated with the measures, analogously to Fig. \ref{fig:shear_1bin}.}}
\label{fig:shear_3bin}
\end{figure}

In Fig. \ref{fig:shear_1bin} we show the tunnel stacked tangential shear profiles for all the analyzed models. Firstly, we note that all tangential shear profiles are statistically negative, consistent with underdensities, and have an amplitude of the order of $10^{-3}$, matching values found in the literature for tunnel voids. In addition, we detect a $\gamma_t(r_p)$ signal that remains negative out to the value of the radius assigned to our voids, resembling the behavior typically observed in WL minima-based finders \citep{Davies2019CosmologicalTestGravityWeakLensing, Davies2021OptimalVoidFinderinWeakLensing, Davies2021ConstrainingCosmoWeakLensingVoids}.
In particular, the observed trends are in agreement with theoretical expectations, namely, there are deeper tangential shear signals for MG models with stronger $|f_{R0}|$ intensities and a shallower signal in the presence of massive neutrinos. This is explained by the fact that cosmic voids experience a depletion of their internal density profiles when their evolution is enhanced; conversely, they are less underdense when their growth is hampered.\\
\indent We note, however, that there is a pivot point at $r_p/R_v \simeq 4$ where all model results almost degenerate. Over this scale, there is an inversion in the trends of the profiles. This point marks the transition between regions influenced by underdensities and those affected by overdensities. Moving toward the outskirts of 2D voids, the tangential shear profiles begin to deviate from $\Lambda$CDM. This deviation follows the expected effects of overdensities, with an increased signal for MG models and a reduced signal for scenarios with massive neutrinos. The transition scale depends on the chosen definition of the 2D void radius. Although we tested several S/N thresholds in the range $[-1, -3]$, we found that $-2.5$ offers the best trade-off between shear signal strength and void sample purity. Increasing the threshold shifts the pivot point to smaller $r/R_v$ values, while lowering it moves the crossing outward. However, no clear predictive trend was observed. Higher thresholds also increase the incidence of 2D void mergers, reducing the void count and affecting the reliability of cosmological constraints. Therefore, using values near zero proved unfeasible.\\
\indent Additionally, we focused on studying the dependency of the stacked tangential shear profile on the void size. This analysis allowed us to examine how the tangential shear properties vary with void radii, while providing valuable insights into the matter distribution around voids at different scales.\\
\\
\indent For each cosmological model, we divided the entire sample of our voids according to their size, into three equipopulated bins:
\begin{itemize}
    \item small voids with radii $R_{\rm v}$ in the range $[1, 3.97] \ \mathrm{arcmin}$;
    
    \item medium voids with radii $R_{\rm v}$ in the range $]3.97, 4.85] \ \mathrm{arcmin}$;
    
    \item large voids with radii $R_{\rm v}$ in the range $]4.85, 15] \ \mathrm{arcmin}$.
\end{itemize}
Then we extracted the void tangential shear profiles from these equipopulated subsamples. The result is represented in Fig. \ref{fig:shear_3bin}.

\begin{figure}[ht]
\centering
\includegraphics[width=\columnwidth]{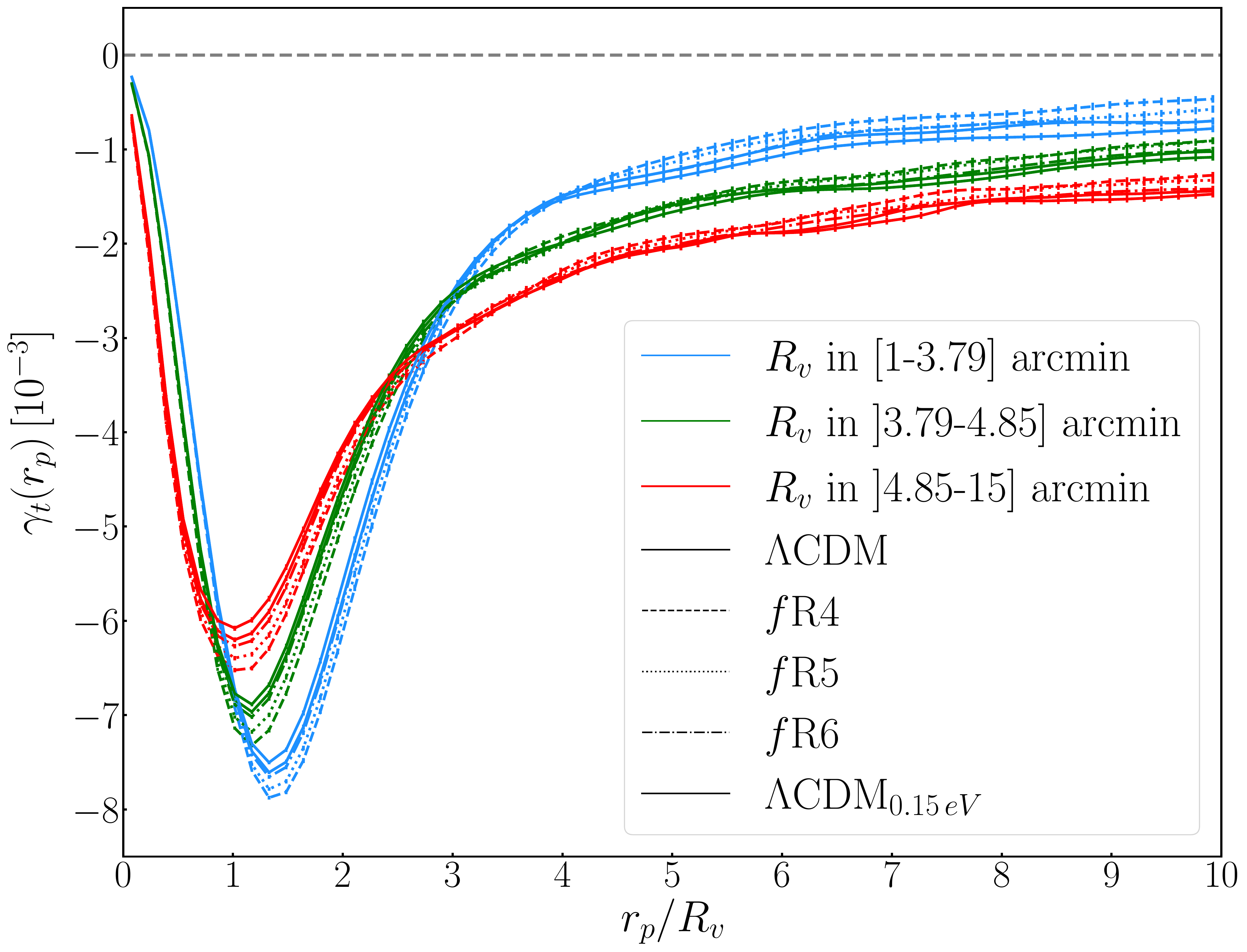}
\caption{\small{Stacked tangential shear profiles of all the analyzed cosmologies, divided into subsamples of voids with different sizes. The split of the void sample is the same as presented in Fig. \ref{fig:shear_3bin} and is used to distinguish voids of small (blue), medium (green), and large (red) sizes. The results for different cosmologies are shown using different line styles, as reported in the legend.}}
\label{fig:shear_all_3bin}
\end{figure}

\indent From the top panel of this figure, we can identify a trend between the three bins. In particular, we note that the signal of small voids is deeper and rises more rapidly towards zero, while for the subsample of large voids, we observe a less profound signal and a slower rise. 
This behavior is similar to the one found for the density profiles of 3D voids \citep{Nadathur2015SelfSimilarityvoidsdensity, Hamaus2014UniversalProfile, Voivodic2020}, for which small voids show, in fact, deeper interiors and high compensation walls. This occurs because large voids have internal substructures and irregular shapes due to merger events. Furthermore, as the VSF analysis shows, small voids are much more numerous than large ones. As a result, the bin for large voids covers a broader range of radii. This causes the stacked signal for large 2D voids to be influenced by the varying tangential shear profiles, leading to a smoother average trend.\\
\indent The bottom panel of Fig. \ref{fig:shear_3bin} shows the residuals for the three cases, computed with respect to the tangential shear profile of the complete sample of voids, $\overline{\gamma_t}(r_p)$, which serves as our reference. We finally observe that the subsample of voids of intermediate size is the most similar to the average tangential shear profile computed with the totality of the void sample.\\
\indent In Fig. \ref{fig:shear_all_3bin}, we present the stacked tangential shear profiles, divided into the three equipopulated bins for each cosmological model. This approach highlights the significant variations in the profile trends across the different models analyzed. It reveals a systematic trend in the amplitude and shape of the tangential shear profiles, with smaller voids (blue) exhibiting deeper signals compared to larger voids (red). This behavior reflects the distinct structural properties of voids at different scales and also in the alternative cosmological models analyzed. Furthermore, this observed dependency in void size and cosmological model will serve as a key parameter in Sect. \ref{Sec:modeling}, where we test different parameterizations to describe the tangential shear signal from WL tunnel voids. We finally highlight that, in the analyzed scenarios, the variations caused by differences in void size are more pronounced than those caused by changes in the cosmological model when considering voids of similar radius.
\section{Modeling tangential shear profile of tunnel voids}\label{Sec:modeling}
In the final part of this work, we aim to model the tunnel void tangential shear profile using a parametric formula. 
Although no theoretical model based on first principles exists for the WL signal from voids, an empirical function can be used to fit the data to analyze the behavior and limits of its free parameters in different scenarios. To achieve this, we perform a Bayesian analysis by running a full Markov chain Monte Carlo (MCMC) procedure to sample the posterior distribution of the free parameters in the considered models. This approach does not directly constrain the cosmological parameters but represents a first step to understanding the physics behind the WL phenomenon in voids in order to test GR and MG models.

\subsection{Tunnel void tangential shear from density profiles}\label{subsec:ESMD_modeling}
The first method we followed is to start from a known parametric formula from the literature to represent the density profile of 3D voids. Following the approach of \cite{Pisani2014}, \cite{Fang2019} and \cite{Boschetti2023}, we can integrate the functional form of the density profile along the line of sight to obtain the projected surface mass density, $\Sigma(r_p)$, which we can then relate to convergence through Eq. \eqref{eq:K_sigma} and  derive the tangential shear profile, $\gamma(r_p)$.

The first functional form we consider to represent the density profile of 3D voids is the Hamaus-Sutter-Wandelt \citep[HSW,][]{Hamaus2014UniversalProfile}. It takes the form 
\begin{equation}\label{HSW}
\frac{\rho_\mathrm{v}(r)}{\bar{\rho}}-1=\delta_{c} \frac{1-\left(r / r_\mathrm{s}\right)^{\alpha}}{1+\left(r / r_\mathrm{v}\right)^{\beta}} \, .
\end{equation}
It is characterized by four free parameters ($\delta_{c}$, $\alpha$, $\beta$, $r_\mathrm{s}$), which represent, respectively, the central density contrast, the inner and outer slopes of the profile, and the characteristic scale where $\rho_\mathrm{v}=\bar{\rho}$. This functional form is commonly used to fit voids of different sizes and internal densities.

The second functional form we consider is the hyperbolic tangent \citep[HT,][]{Voivodic2020}. It has only two free parameters, $\delta_c$ and $s$, and is expressed as:
\begin{equation}\label{TANH}
\rho_{v}\left(r/R_{\rm v}\right) = 1 + |\delta_c| \, \bar{\rho} \left\{ \frac{1}{2}\left[1+\tanh \left(\frac{y-y_{0}}{s\left(R_{\rm v}\right)}\right)\right] -1 \right\} \, .
\end{equation}
In this parametrization, $\delta_c$ is the density contrast at the void center, $y=\ln \left(r / R_{\rm v}\right)$ and $y_{0}=\ln \left(r_{0} / r_{v}\right)$. The radius $r_{0}$ is determined by requiring that the integral of the profile up to $R_{\rm v}$ be $\Delta_{v}=0.2$. This allows us to express $r_{0}(s)$ (in units of $h^{-1} \ \mathrm{Mpc}$ ) as a second-order polynomial function: $r_{0}(s)=0.37 s^{2}+0.25 s+0.89$, where $s$ represents the gradient of the profile. The parameter $s$ works similarly to the concentration parameter in the NFW profile, governing the rate at which density increases as we move away from the center of the void.

As shown in Fig. \ref{fig:Maggiore}, neither of these functions leads to a good fit of our data. Despite the large degrees of freedom of the considered parametric forms and the very large prior assigned to their coefficients, the complex shape of the tangential shear profiles could not be reproduced accurately. For both cases, the reduced $\chi^2$ obtained with the best-fit parameters is indeed very far from unity, namely, $\tilde{\chi}_{\rm HSW}^2 \simeq 2031$ and $\tilde{\chi}_\mathrm{HT}^2 \simeq 7750$.

The motivation for this inconsistency must be sought from the very nature of our voids. In our case, we are trying to model the WL signal generated by tunnel voids, which are the result of projecting numerous underdensities along the line of sight. Following the approach described in this section, we are instead imposing the modeling of the tangential shear through the projection of the typical density profile of isolated 3D voids. This assumption cannot be valid for our tunnel voids, which instead derive from the projection of a complex distribution of underdensities with different sizes and positions. Therefore, we emphasize that in the analyses of \cite{Fang2019} and \cite{Boschetti2023}, the usage of the 3D density profiles was effective for the following reasons. Both authors make use of a tomographic approach, considering only the WL signal generated by the matter distribution present in relatively thin redshift slices. In this way, they exploit a lens plane that follows the simplified condition of having only one, or in any case a few, 3D voids along the line of sight.
In a future study, we will analyze the connection between the average tangential shear of individual 3D voids and that of tunnel voids identified with our finder using a tomographic approach.

\subsection{A new parametric formula}\label{subsec:new_model}
Here, we present and validate a new parametric function suitable for the modeling of the tunnel voids' tangential shear profiles extracted from our void catalogs.
Our goal is to accurately reproduce the shape and amplitude of our shear profiles, employing the smallest possible number of free parameters. Moreover, we want this new functional form to be flexible enough to represent the shear profiles both for different void sizes and the five cosmological models analyzed in this work. A useful example to understand the degree of ``flexibility'' we need to reach with this function is shown in Fig. \ref{fig:shear_all_3bin}. To the best of the authors’ knowledge, this is the first and so far only parametric formula proposed in the literature specifically developed to model the tangential shear signal of WL tunnel voids.

The functional form found that best met these requirements can be expressed as
\begin{equation}\label{eq:newparametric}
     \gamma_t(r_p) = a \cdot r_p \left( \frac{1 - r_p^b}{1 + r_p^c} - \frac{\exp(d \cdot r_p)}{1 + \exp(e \cdot r_p - (d+e))} \right) \ .
\end{equation}
The coefficients $a$, $b$, $c$, $d$, and $e$ represent the five free parameters of our model. For now, these parameters are not intended to have a physical meaning but rather to regulate the shape and amplitude of the profile across different scales. Specifically, the parameter $a$ is the normalization of the curve, while $b$ shifts the position of its minimum. The steepness of the rise in the outer part of the profile is modulated by $c$, whereas $d$ controls how deep the profile declines at its minimum. Finally, $e$ defines the scale at which the exponential growth begins to dominate. To understand in detail their role, in Appendix~\ref{app:5_param}, we illustrate and discuss the effect that varying singularly each parameter produces on the shear profile.
\begin{figure}
\centering
\includegraphics[width=\columnwidth]{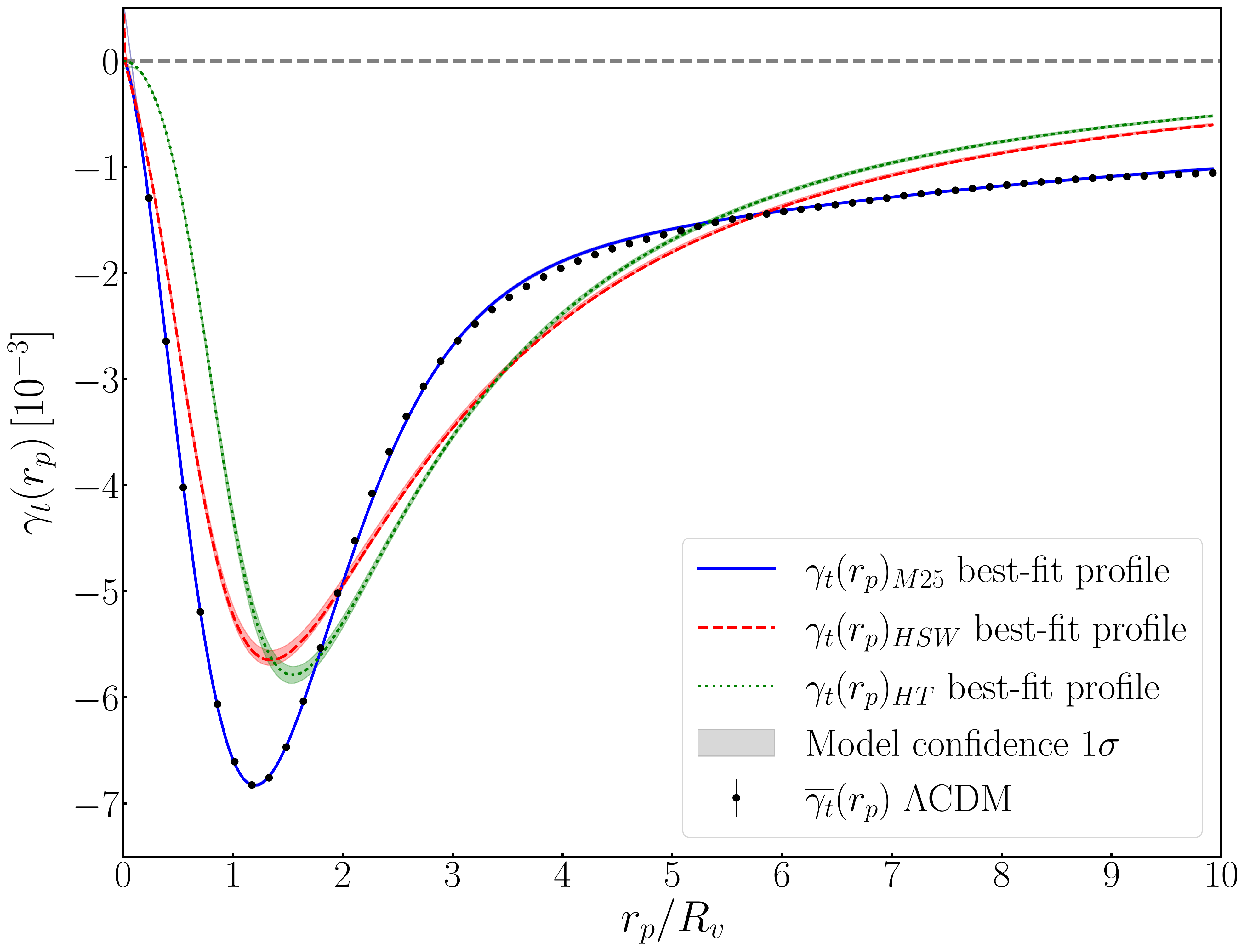}
\caption{\small{Best fits of the averaged stacked tunnel voids tangential shear profiles, $\gamma_t(r_p)$, measured in $\Lambda$CDM cosmology adopting three different parametric functions. The data are derived as the mean of the $256$ $\Lambda$CDM mock light cones and represented with black markers having error bars computed as specified in Sect. \ref{subsec:WL}. The first two models considered are derived from the line-of-sight integration of two 3D void density profile functions: HSW (red dashed line) and HT (green dotted line). The last one is the new formula developed in this work for WL tunnel voids (blue solid line), reported in Eq. \eqref{eq:newparametric}. The shaded area, in some cases barely visible, represents the $1\sigma$ uncertainty associated with each fit.}}
\label{fig:Maggiore}
\end{figure}
\begin{figure*}
\centering
\includegraphics[width=0.8\textwidth]{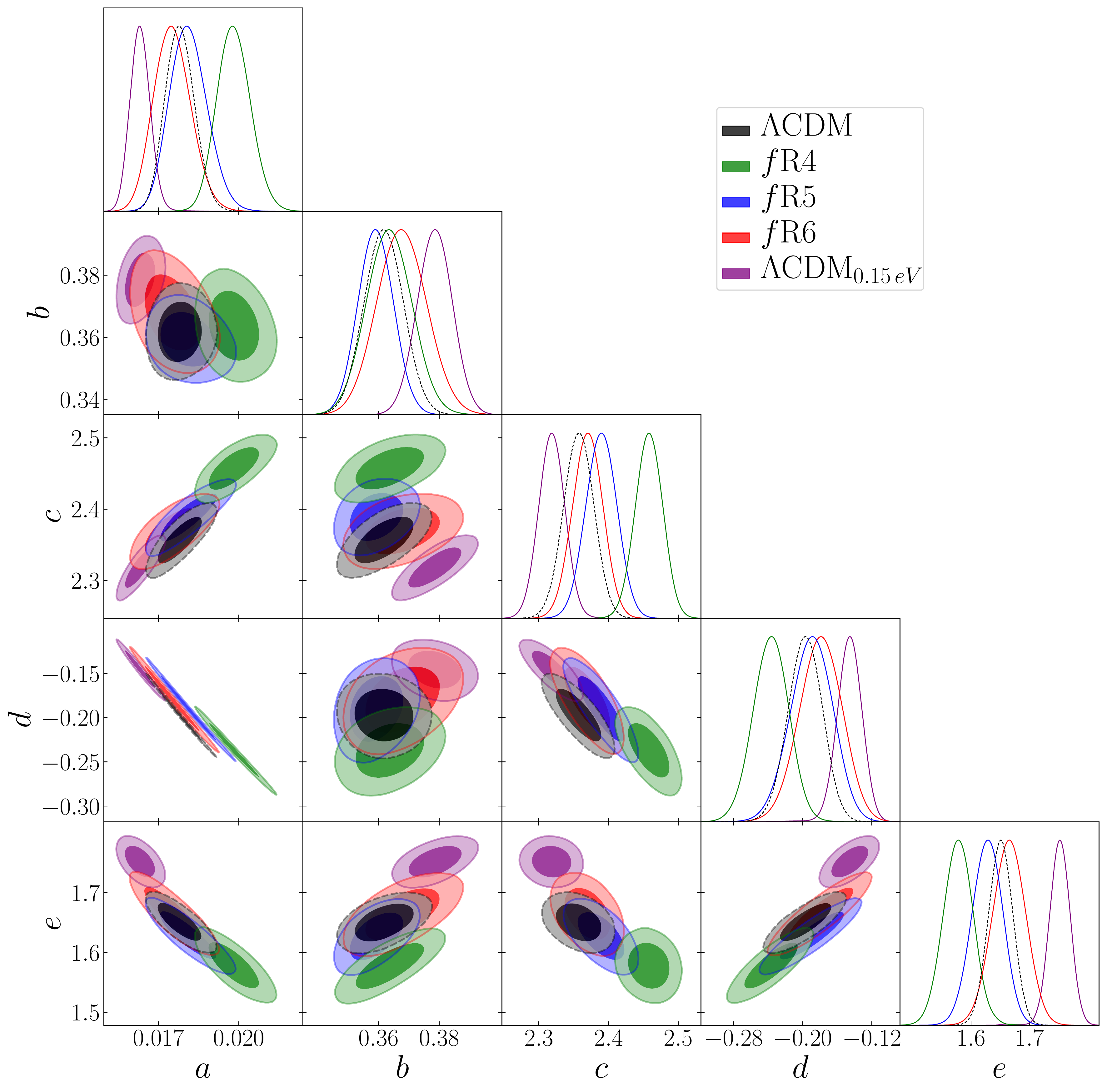}
\caption{\small{$68\%$ and $95\%$ confidence levels for the five parameters of the functional form reported in Eq. \eqref{eq:newparametric}, used to model the tunnel void tangential shear profiles extracted from simulations featuring different cosmological scenarios: $\Lambda$CDM (black), $fR4$ (green), $fR5$ (blue), $fR6$ (red), $\Lambda$CDM$_{0.15 \, \mathrm{eV}}$ (purple).}}
\label{fig:contours_cosmo}
\end{figure*}

In Fig. \ref{fig:Maggiore}, we present one of the most important results of this work, namely the application of the new function for fitting the tunnel voids stacked tangential shear profile extracted from the $\Lambda$CDM simulations. It is easy to notice that there is an excellent match with the data across all scales. The reduced $\chi^2$ is indeed now closer to the unity, namely, $\tilde{\chi}_{\rm M25}^2 \simeq 0.37$. The low chi-squared value might suggest that the model could have more free parameters than necessary. However, it is important to note that the same model will also be used to fit the shear profile for voids selected by size and in MG scenarios, which shows a large and complex variation at all scales analyzed. Moreover, the reduced $\chi^2$ obtained neglecting the off-diagonal terms is $\tilde{\chi}_{\rm M25}^2 \simeq 4.44$, indicating a strong correlation between the radial bins (see Appendix \ref{app:correlation} for details).\\
\indent As previously shown in Fig. \ref{fig:shear_3bin}, there is a high dependency of the tangential shear profiles on the void size. According to the selected void radius range, the depth and the position of the profile minimum change, as well as the slope and the height of the outer part. The parametric formula proposed in Eq. \eqref{eq:newparametric} was built with enough degrees of freedom to capture these variations. This is analyzed in Appendix \ref{app:void_size_dependency}, where we provide the best-fit values of the coefficients for the $\Lambda$CDM scenario, splitting the void sample into the same size bins shown in Fig. \ref{fig:shear_3bin}.\\
\indent Now, we want to focus on the cosmological discriminating power of the tunnel void tangential shear profiles, and to do this, we will examine the posterior distribution of the parameters of the function in Eq. \eqref{eq:newparametric}. Notice that, in this analysis, firstly we make no distinctions based on the size of the voids and perform the stacking of all the voids extracted for a given cosmology.\\
\begin{figure*}
\centering
\includegraphics[width=0.8\textwidth]{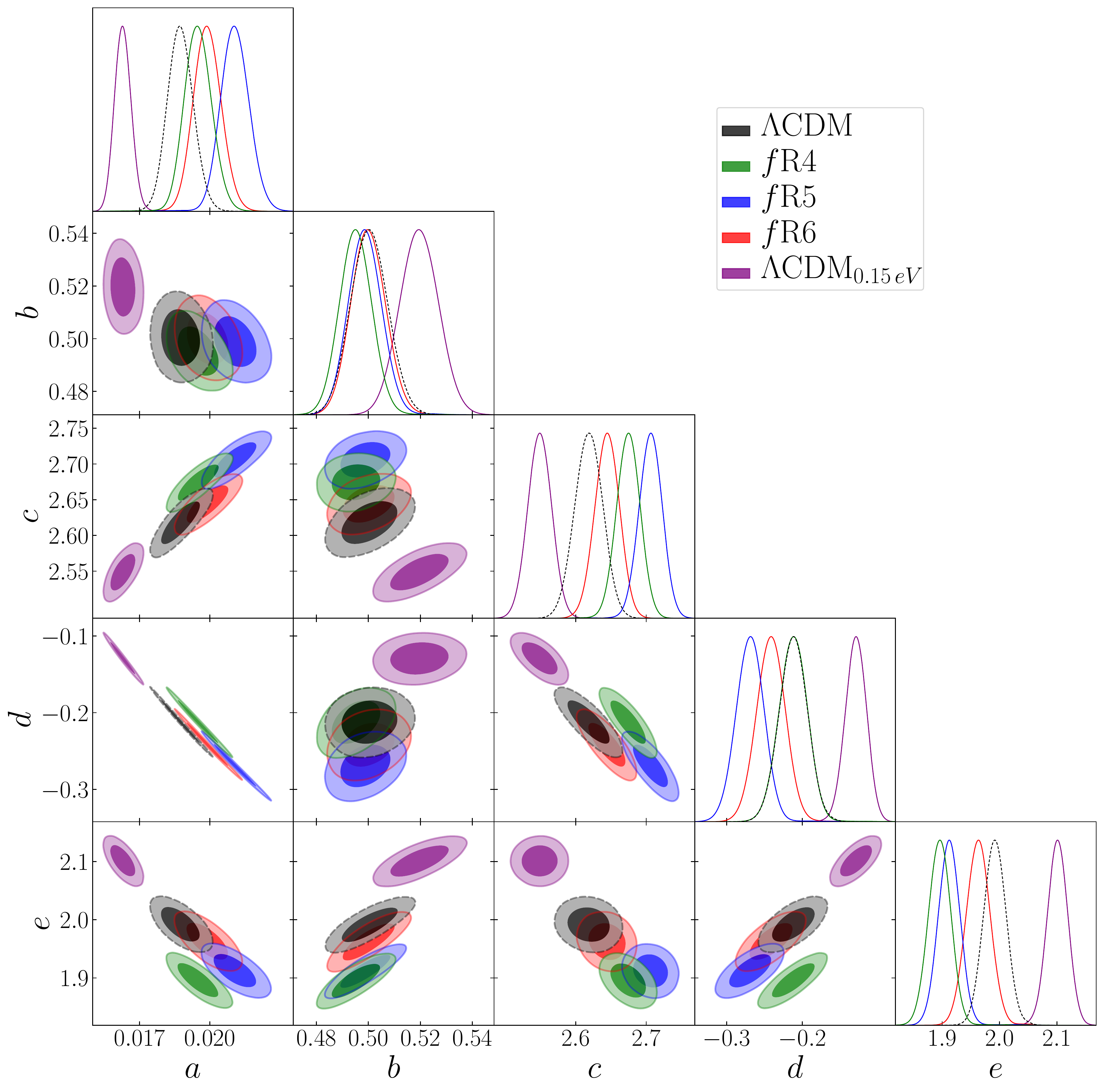}
\caption{\small{Same as Fig. \ref{fig:contours_cosmo}, but for a selection of voids with radius in the range $]3.79, 4.85]$ arcmin.}}
\label{fig:contours_cosmo_medi}
\end{figure*}
\indent We report in Fig. \ref{fig:contours_cosmo} the confidence contours obtained from fitting the shear profiles extracted in simulations featuring the five cosmological models considered in this work: $\Lambda$CDM, $fR4$, $fR5$, $fR6$, $\Lambda$CDM$_{0.15 \, \mathrm{eV}}$. Overall, we note that the confidence contours are mostly ordered according to the level of enhancement or damping of the growth of LSS, with $fR4$ and $\Lambda$CDM$_{0.15 \, \mathrm{eV}}$ showing the largest separation from the $\Lambda$CDM case, but in opposite directions. We note, however, that the confidence contours result in partially overlapping, especially when considering the models characterized by the smaller deviations from the $\Lambda$CDM case. The area of the confidence contours is in fact significantly enlarged by the correlations between radial bins, which decrease the statistical disentangling between the scenarios. Nevertheless, the separation between the most extreme cases -- $fR4$ and $\Lambda$CDM$_{0.15 \, \mathrm{eV}}$ -- from the reference $\Lambda$CDM is over 1$\sigma$ for all the parameters of the model.\\
\indent It is also apparent from this plot that notable positive correlations are present in the $(a, c)$ and $(d, e)$ planes, while negative correlations are observed in the $(a, e)$ and $(a, d)$ ones. The latter, in particular, is characterized by tilted and very elongated ellipses, indicating a strong correlation between these parameters. We refer to Appendix \ref{app:5_param} for a deeper understanding of the behavior of the model coefficients. In Fig. \ref{fig:contours_cosmo_medi}, we present the same confidence contours derived for all analyzed cosmological models, but restricted to the sub-sample of tunnel voids with $R_{\rm v}$ in the range $]3.79, 4.85]$ arcmin.\\
\indent The effect of the void radius selection is not very strong, showing the robustness of the methodology. However, we note that a particular selection in size may help the distinction between specific scenarios. For example, the constraints computed for the $\Lambda$CDM$_{0.15 \, \mathrm{eV}}$ case appear now more detached from the reference $\Lambda$CDM. This might suggest that selecting certain spatial scales could enhance the effects caused by the presence of massive neutrinos, making it more favorable for studying cosmological scenarios with these characteristics. Additionally, selecting by radius prevents the averaging of very different profile behaviors. This implies that WL analyses of voids can be further optimized by carefully selecting void size ranges to enhance sensitivity to both modifications of gravity and neutrino mass effects.\\
\indent For the sake of conciseness, we do not show the results for the other two radius selections (i.e., small and large voids). However, we clarify that the results are consistent with those presented for medium-sized voids. Finally, we emphasize once again that these filters in void size can be used to leverage specific differences between the analyzed cosmological models. For example, we found that applying the selection for large voids ($]4.85, 15]$ arcmin) makes it possible to highlight the differences between the $fR6$ and $\Lambda$CDM cases. \\
\indent We finally underline that the confidence contours presented in this section are meant to be representative of a photometric survey like the one currently being carried out by \textit{Euclid} (see Sect. \ref{subsec:LC} and \ref{subsec:maps}). In our case, however, each of our $256$ maps has a sky area of $25$ deg$^2$. Such a configuration can be interpreted as a mosaic of $256$ semi-independent observations, leading to an effective sky area not greater than $6400$ deg$^2$, which is more than a factor  of $2$ smaller than the one planned for the final data release of \textit{Euclid} (i.e., $14000$ deg$^2$). We believe, therefore, that the constraining power showcased by this analysis to be close to one half ($45.7\%$) of the one expected for the \textit{Euclid} wide survey.\\
\indent To summarize this section, we can conclude that our identified tunnel voids are sensitive to the effects of MG and massive neutrinos. Furthermore, the modeling of their tangential shear profiles via the newly proposed parametric formula can offer new insights to analyze the variations induced by these cosmological models.

\section{Summary and conclusions}\label{sec:conclusions}
In this work, we investigated the WL signal of tunnel voids in the convergence field, focusing on MG models within the $f(R)$ framework \citep{HuSawiki2007FR} and their effects with respect to those of massive neutrinos. These alternative $f(R)$ models are considered viable as they reproduce both early- and late-time cosmic acceleration without a cosmological constant, while employing screening mechanisms to remain consistent with GR on small scales and in high-density regions. We also included a massive neutrino model to disentangle its degenerate effects from those of MG on LSS formation.\\
\indent Previous studies have demonstrated that cosmic voids are powerful cosmological probes for testing deviations from GR, particularly in the context of $f(R)$ models. For instance, 3D void statistics have been used to constrain $f(R)$ gravity and massive neutrino scenarios \citep{Contarini2021_voids_MG_neutrinos}, demonstrating their ability to break degeneracies between these components. Similarly, WL statistics of both 2D and tunnel voids have exhibited enhanced sensitivity to $f(R)$-induced modifications of structure formation in lensing surveys similar to \textit{Euclid} and LSST \citep{Cautun2018}.\\
\indent Using an innovative pipeline, we analyzed the S/N of noised and smoothed convergence maps from the \dustp $N$-body simulations \citep{Giocolietal2018_mio, Peel2018}, systematically comparing their behavior with the standard $\Lambda$CDM model. We summarize our main results below. 

\begin{itemize}
\item We presented a 2D WL tunnel void finder called \texttt{PyTwinPeaks} that allows computationally efficient identification of underdense regions in convergence maps, providing precise measurements of void centers, sizes, and geometries without requiring 3D galaxy positions. 

\item We studied the void size function, which revealed that models with stronger MG effects display an increased number of tunnels, while the presence of massive neutrinos suppresses their abundance.

\item We analyzed the stacked tangential shear profile of identified tunnel voids. We extracted the average stacked tangential shear profile across the five cosmological models, finding that voids in $f(R)$ MG scenarios exhibit a more pronounced shear signal owing to their deeper density contrasts. Conversely, voids in $\Lambda$CDM$_{0.15 \, \mathrm{eV}}$ display a weaker signal due to the suppression of LSS growth.

\item We investigated the behavior of the tangential shear profile according to the size of voids using three equipopulated subsamples. Smaller voids show a deeper and steeper signal, while larger voids have a shallower and flatter signal.

\item We modeled the void shear profiles, developing a new five-parameter empirical parametric formula. Importantly, it  fits all the stacked shear profiles accurately, making it a valuable tool for analyzing WL signals in tunnel voids.
\end{itemize}
\indent It is important to underline that the differences in the shear measurements can be used to discriminate between various cosmological models, revealing the presence of MG effects. This is especially important at the minimum of the profile, where this approach enables a clear statistical separation. A Bayesian analysis was performed to constrain the free parameters of the proposed formula. We first applied MCMC sampling to the $\Lambda$CDM shear profiles, analyzing the corresponding confidence contours to examine parameter correlations. Scrutinizing different cosmologies, we observed that confidence contours shift according to structure growth rates, with $fR4$ and $\Lambda$CDM$_{0.15 \, \mathrm{eV}}$ displaying the most extreme behavior.\\
\indent Given our encouraging findings, we plan to further expand this project in several ways. First, the current version of \texttt{PyTwinPeaks}, which we have already made publicly available, will be restructured to handle both WL underdensities and overdensities. To strengthen the statistical robustness of our results, we also plan to apply the pipeline to larger-volume simulations \citep{CastanderFS2} and explore a broader range of cosmological models \citep{EC_higherorderWL_Ajani2023}.\\
\indent This development is supported by the modular structure of our pipeline, where each component can be independently adapted, making the methodology applicable across different cosmological suites of mock catalogs. The core of the method is the 2D void finder, which can be easily configured to process input maps produced with various lensing reconstruction techniques, including full-sky multiplane ray-tracing or tomographic approaches. Additionally, shear-based lensing estimators can replace convergence to improve compatibility with real data affected by masking or anisotropic noise in $\kappa$ reconstruction. Moreover, the new formula for WL tunnel voids is expected to generalize across a wide range of scenarios.\\
\indent An extended WL tomographic analysis will be performed to track signal variations across redshifts and identify the optimal range for studying void tangential shear profiles. By following the evolution of cosmic structures in redshift bins, we aim to extract higher-order WL statistics that capture key aspects of structure formation and growth. This will support forecasts for Stage-IV LSS surveys and enable tests of dynamical dark energy and MG. Ultimately, combining WL statistics from both voids and clusters will yield tight cosmological constraints and offer a unified framework for probing fundamental questions in cosmology.\\
\indent Furthermore, we will explore the feasibility of applying our analysis to real data catalogs. Our new tunnel void pipeline will be applied to the first data release of the ESA-\textit{Euclid} mission, allowing for a direct observational test of our methodology. The integration of standard cosmological probes alongside WL measurements from voids will maximize the information extracted, leveraging the complementarity of constraints from under- and overdensities of LSS \citep{Bayer2021, Kreisch2022, Contarini2023_param, Pelliciari2023}. The photometric galaxy catalog from \textit{Euclid} will allow us also to analyze the cross-correlation between WL and underdense regions, namely, the void-lensing cross-correlation \citep[see e.g.,][]{Bonicietal2023}.\\
\indent Finally, our long-term objective is to develop a theoretical model for the tangential shear profile of cosmic voids from first principles, a step that remains an open challenge. This will require exploring semi-analytical models to provide a more comprehensive framework for using WL tunnel voids and 3D void lensing as a cosmological probe. By deriving a physically motivated model, we aim to directly constrain cosmological parameters and their evolution over time, improving the predictive power of void-lensing studies.

\section*{Data availability}\label{sec:data_link}
\addcontentsline{toc}{section}{Data availability}
To our knowledge, no 2D void finder has been publicly released to date. To support reproducibility and further development, we provide our WL tunnel void finder implementation as open-source code at \url{https://github.com/LeonardoMaggiore/PyTwinPeaks}.\\
\\
\begin{acknowledgements}
We thank the anonymous referee for his/her comments that helped improve the presentation of our results. We would like to thank A. Pisani, N. Hamaus, N. Schuster, and C.T. Davies for their help, useful discussions, and fruitful collaborations. We also acknowledge the use of computational resources from the parallel computing cluster of the Open Physics Hub (\url{https://site.unibo.it/openphysicshub/en}) at the Physics and Astronomy Department in Bologna. LM and CG acknowledge the financial contribution from the PRIN-MUR
2022 20227RNLY3 grant 'The concordance cosmological model:
stress-tests with galaxy clusters' supported by Next Generation EU and from the grant ASI n. 2024-10-HH.0 “Attività scientifiche per la missione Euclid – fase E”. GC thanks the support from INAF theory Grant 2022: Illuminating Dark Matter using Weak Lensing by Cluster Satellites. We acknowledge the use of the Python libraries \textsc{NumPy} \citep{Harris2020}, \textsc{Matplotlib} \citep{Hunter2007}, \textsc{Cobaya} \citep{Torrado2021_cobaya} and \textsc{GetDist} \citep{Lewis2019_GetDist}.
\end{acknowledgements}

\bibliographystyle{aa}
\small
\bibliography{bibliography}
\normalsize
\appendix

\section{Covariance matrix}\label{app:correlation}
In this appendix, we present the tangential shear covariance matrix for the WL tunnel void sample of $\Lambda$CDM scenario. The diagonal of this matrix is used to calculate the errors in the tunnel voids stacked tangential shear profile through Eq. \eqref{eq:gt_errors}, while its inverse is inserted in the likelihood to compute the fit of the data through Eq. \eqref{eq:newparametric}.

\begin{figure}
    \centering
    \includegraphics[width=\columnwidth]{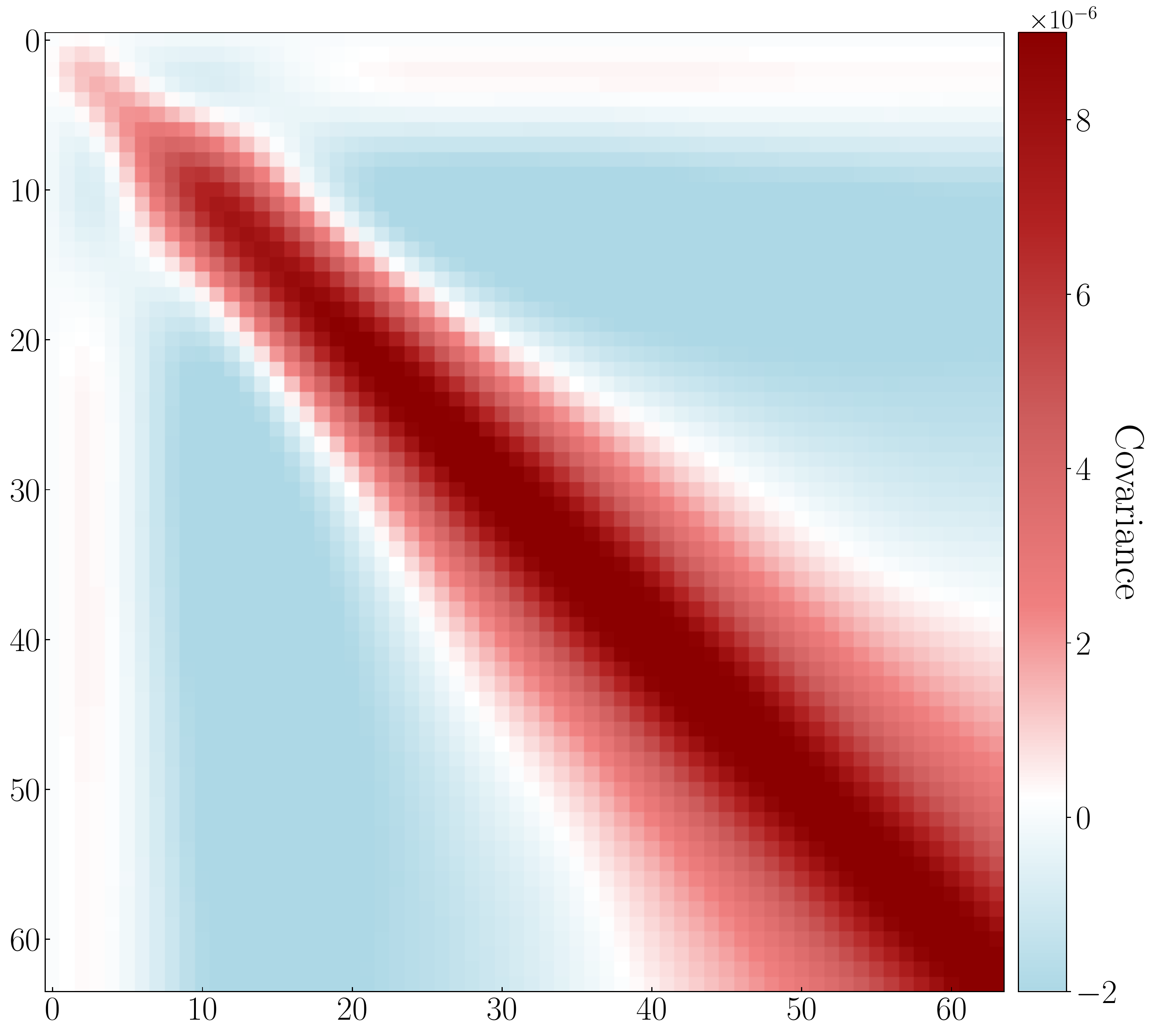}
    \caption{Tunnel voids tangential shear profiles covariance matrix measured for the entire sample of voids in the $\Lambda$CDM scenario, in each of the $64$ radial bins.}
    \label{fig:cov_mat}
\end{figure}

Figure \ref{fig:cov_mat} shows the measured covariance matrix, $C_{ij}$, for WL tunnel voids identified through the developed finder, in WL maps with GSN included and with a smoothing scale of $\theta_{\rm{s}}=2.5$ arcmin. It is computed via the formula
\begin{equation}\label{eq:cov_mat}
    C_{ij} = \frac{1}{N-1} \sum_{k=1}^{N} [\gamma_{\rm{t}}(i) - \bar{\gamma}_{\rm{t}}(i)][\gamma_{\rm{t}}(j) - \bar{\gamma}_{\rm{t}}(j)] \, ,
\end{equation}
where $i$ and $j$ are radial bin indices and $N = 22203$ is the total number of samples of WL tunnel voids in the $\Lambda$CDM scenario. $\gamma_{\rm{t}}$ is the tangential shear and an over-bar denotes the mean from $N$ WL tunnel voids. Since we considered our light cones to be independent realizations, $N$ is not the number of our maps (as in \citealp[]{Davies2021OptimalVoidFinderinWeakLensing}) but the number of voids used to calculate the covariance matrix.
The size of this matrix is of $64 \times 64$ and represents the statistical relationship between the radial bins used to calculate the stacked tangential shear profile.

For the calculation of the log-likelihood used in the Bayesian analysis presented in Sect.~\ref{subsec:new_model}, we used the precision matrix obtained by inverting the tangential shear covariance matrix. This inverse was corrected using the Anderson-Hartlap factor \citep{Anderson2003, Hartlap2007}, $\alpha$, through the relation $C_{\text{corr}}^{-1} = \alpha C^{-1}$, where $C^{-1}$ is the inverse of the sample covariance matrix. The $\alpha$ factor is given by
\begin{equation}\label{eq:alpha_Hartlap}
\alpha = \frac{ N - N_{\rm{bin}} - 2 }{N - 1} \, ,
\end{equation}
where $N$ is the total number of voids in the sample, and $N_{\rm{bin}} = 64$ is the number of radial bins. We use this method\footnote{We emphasize that to verify the accuracy of this approach, we also computed the errors on the tangential shear profile using jackknife and bootstrap techniques, obtaining results with the same order of magnitude. This is due to the high statistics of the identified WL tunnel voids.} to correct for the bias introduced by inverting a covariance matrix of profiles extracted from noised maps. 

The covariance matrix shown in Fig.~\ref{fig:cov_mat} exhibits a dominant diagonal structure, indicating a strong correlation between the adjacent bins. This effect is caused by the overlap between the void shear profiles of the same S/N map, which occurs especially at larger distances from the center. Covariance matrices computed from alternative cosmologies, while remaining of the same order of magnitude, exhibit variations at the $10^{-6}$ level relative to the $\Lambda$CDM case, reflecting changes in LSS.

The smooth gradient observed in the off-diagonal elements highlights the progressive nature of correlations, with stronger covariance at large separations. This suggests that the uncertainty in the tangential shear profile does not behave as purely uncorrelated Gaussian noise but instead encodes features intrinsic to the void environment. Additionally, the overall amplitude of the covariance matrix is a direct indication of the sample variance in the WL tunnels measurement, implying that increasing the number of detected voids would reduce uncertainties proportionally.

A key observation is that correlations remain relatively strong for large radial distances, where the signal transitions into the background noise. The transition from positive to negative correlations in off-diagonal elements also suggests complex interactions between shear contributions from neighboring voids, which could be further investigated in future studies.

Overall, these matrices provide critical insights into the statistical properties of the measured WL tunnel void signal. The observed correlations highlight the necessity of including the full covariance structure in parameter estimation and model fitting, as neglecting these correlations could lead to biased constraints on cosmological parameters. Future improvements in the void detection algorithm and larger survey volumes, such as those expected from \textit{Euclid}, will be crucial in reducing the sample variance and enhancing the robustness of WL tunnel voids and overdense structures as a cosmological probe.

\section{Free parameters of the model}\label{app:5_param}
\begin{figure*}
\centering
\includegraphics[width=\textwidth, trim={0 5.5cm 0 5.5cm}, clip]{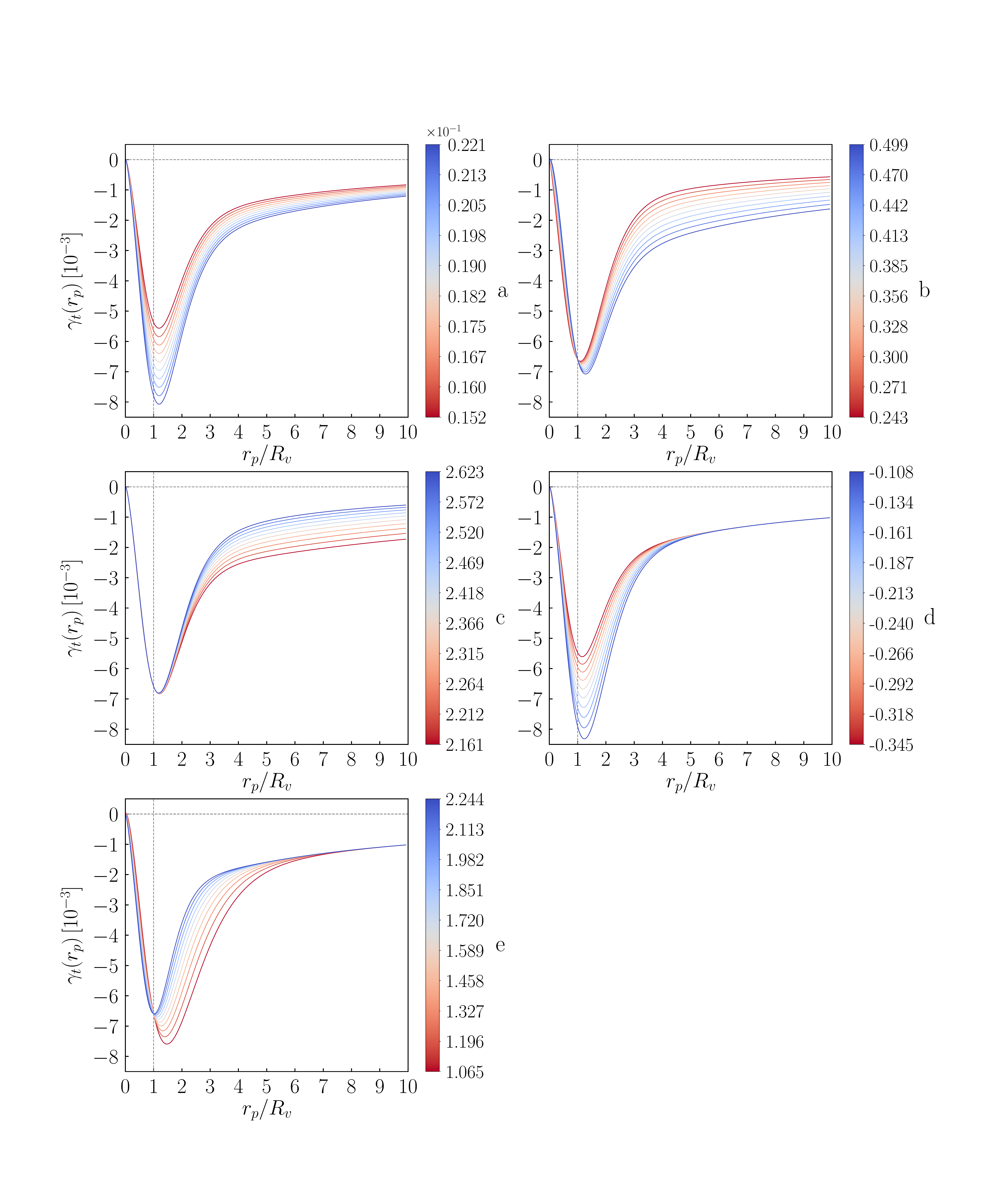}
\caption{\small{Effects of varying individually the coefficients of the function reported in Eq. \eqref{eq:newparametric}. For each parameter, ten curves are represented and colored according to the value of the selected parameter. The colorbar located on the right of each subplot shows the association of the color used with the parameter value. The different panels refer to the parameters $a$ (top left), $b$ (top right), $c$ (central left), $d$ (central right), and $e$ (bottom left).}}
\label{fig:block}
\end{figure*}

Here, we focus on the interpretation of the coefficients of Eq. \eqref{eq:newparametric}, providing a brief description of the main effects of each parameter:
\begin{enumerate}[label=\alph*), align=left]
    \item Amplitude or normalization. It modifies the amplitude of the function, i.e., its normalization, and it is related to the overall depth of the tunnel void.
    \item Position of the minimum. It plays a role in determining the position of the absolute minimum of the profile, also influencing the amplitude of the profile at large distances from the void center.
    \item Exponential growth. It regulates the exponential rise after the minimum, so the slope of the outer part of the profile.
    \item Depth of the minimum. It determines the minimum value of the profile, and it is related to the central density contrast of the tunnel void.
    \item Starting point of the exponential term. It influences the scale at which the exponential part of the function begins to become dominant.
\end{enumerate}
Figure \ref{fig:block} illustrates the influence of each parameter on the tangential shear profile. Variations in parameter $a$ primarily rescale the entire function, while parameter $b$ shifts the location of the minimum. Parameters $c$ and $e$ control the steepness of the transition at large scales, whereas $d$ directly affects the depth of the void-induced shear signal.\\
\indent Regarding the physical meaning of the parameters, we can currently only hypothesize that it is linked to the complex morphology and density profile of tunnel voids, particularly around the peak of the lensing kernel that marks the segment of the line of sight where the density distribution carries the most weight. (see Sect. \ref{subsec:wlform}).\\
\indent The interplay between these parameters provides a novel approach to characterizing cosmic voids in WL maps. By fitting void shear profiles across different cosmological models, we can extract information on the underlying physics governing LSS formation and evolution. The response of these parameters to varying gravitational models and neutrino masses enables a deeper understanding of how modifications to gravity and the presence of hot dark matter affect the large-scale distribution of mass in the Universe.

The results obtained through this parametric modeling demonstrate that WL tunnel voids offer a robust tool for probing the nature of gravity on cosmic scales. By refining the interpretation of these parameters and linking them to fundamental cosmological quantities, future analyses can enhance the constraining power of void lensing studies and strengthen the role of cosmic voids as a key probe for testing deviations from General Relativity. Extensions of this analysis could explore whether these constraints tighten further with increased void statistics or if systematic uncertainties, such as shape noise or projection effects, influence the parameter degeneracies.

\section{Void size dependency}\label{app:void_size_dependency}

Here we focus on the dependence of the best-fit parameters of Eq. \eqref{fig:Maggiore} on the size of the tunnel voids. The data used for the fit are the averaged void shear profiles extracted from the 256 realizations of the $\Lambda$CDM S/N maps, splitting the void sample into the three equipopulated subsamples of voids described in Sect. \ref{sec:statistics}.
We report in Table \ref{tab:best_fit_params} the best-fit values of each of the five free parameters of the model, together with their associated $1\sigma$ uncertainty. The best fit of our model is obtained by extracting the median value of the individual parameter distributions, which are well approximated by a Gaussian.

In this case, we decided not to represent the confidence contours in a corner plot because the size of these contours would have been too small. In fact, the error associated with these best fits is, in most cases, between $0.4\%$ and $2.7\%$, with an average relative uncertainty around $2\%$. The only exception is the parameter $d$, which for the medium-size tunnel voids shows a relative uncertainty of approximately $9.5\%$. The parameter $c$ appears instead to be the most constrained one, with a percentage error ranging from $0.43\%$ to $0.78\%$. We believe that the parameter values shown in Table \ref{tab:best_fit_params} can be used as a reference for future studies on tunnel voids.

The degeneracy between these parameters is also dependent on the void size, although to an almost negligible extent. We refer to Figs. \ref{fig:contours_cosmo} and \ref{fig:contours_cosmo_medi} for a more in-depth study of the covariance between these parameters with different void size selections.

\renewcommand{\arraystretch}{1.5}
\begin{table}[H]
    \centering
    \caption{$\Lambda$CDM best-fit parameters for voids of different sizes}
    \begin{tabular}{lccc}
        \hline
         & $[1, 3.97]$ & $]3.97, 4.85]$ & $]4.85, 15]$ \\
        \hline
        \(a\) & $0.0085 \pm 0.0002$ & $0.0187 \pm 0.0005$ & $0.0367 \pm 0.0007$ \\
        \(b\) & $0.578 \pm 0.008$ & $0.501 \pm 0.007$ & $0.246 \pm 0.006$ \\
        \(c\) & $2.55 \pm 0.02$ & $2.62 \pm 0.02$ & $2.31 \pm 0.01$ \\
        \(d\) & $0.31 \pm 0.01$ & $-0.21 \pm 0.02$ & $-0.67 \pm 0.01$ \\
        \(e\) & $2.09 \pm 0.01$ & $1.99 \pm 0.02$ & $1.62 \pm 0.04$ \\
        \hline
    \end{tabular}
    \tablefoot{Best-fit values of the five free parameters ($a, \ b, \ c, \ d, \ e$) of Eq. \eqref{fig:Maggiore}, obtained for the $\Lambda$CDM void shear profiles. Each column corresponds to a different selection for void sizes, expressed in arcminutes and increasing from the smallest to the largest, going from left to right. The reported uncertainties represent the $1\sigma$ errors.}
    \label{tab:best_fit_params}
\end{table}

\phantomsection
\label{LastPageAnchor}
\newcounter{LastPageCount}
\setcounter{LastPageCount}{\value{page}}

\AtEndDocument{%
  \immediate\write\@auxout{%
    \string\newlabel{LastPageCountValue}{{}{\theLastPageCount}}%
  }%
}
\end{document}